\documentclass[fleqn,usenatbib]{mnras}
\usepackage{natbib}

\usepackage{newtxtext,newtxmath}

\usepackage[T1]{fontenc}
\usepackage{ae,aecompl}

\usepackage{graphicx}	% Including figure files
\usepackage{amsmath}	% Advanced maths commands
\usepackage{amssymb}	% Extra maths symbols

\usepackage[normalem]{ulem}
\usepackage{xcolor}
\usepackage{nicefrac}

\definecolor{new_color}{rgb}{1.0, 0.43, 0.3}
\newcommand{\add}[1]{\textcolor{red}{#1}}

\newcommand{\obn}[1]{\textcolor{cyan}{[O.B. Note: #1]}}
\newcommand{\ob}[1]{\textcolor{teal}{[O.B. #1]}}
\newcommand{\gb}[1]{\textcolor{magenta}{\bf{[G.B. #1]}}}
\newcommand{\unit}[1]{\boldsymbol{\rm{\hat{#1}}}}
\newcommand{\vect}[1]{\boldsymbol{\rm{#1}}}
\newcommand{\obs}{_{\rm obs}}
\newcommand{\uum}{_{_{\rm m}}}
\newcommand{\uuL}{_{_{\rm L}}}
\newcommand{\uuEATS}{_{_{\rm EATS}}}
\newcommand{\uuPOL}{_{_{\rm POL}}}
\usepackage{caption}
\usepackage{subcaption}
\usepackage{comment}
\title[Polarized GRB Afterglows]{Modeling the Linear Polarization of GRB Afterglows Across the Electromagnetic Spectrum}

\author[Birenbaum \& Bromberg]{
Gal Birenbaum,$^{1}$\thanks{E-mail: galbirenbaum@mail.tau.ac.il}
Omer Bromberg$^{1}$
\\
$^{1}$The Raymond and Beverly Sackler School of Physics and Astronomy, Tel Aviv University, Tel Aviv 69978, Israel\\
}

\date{Accepted XXX. Received YYY; in original form ZZZ}

\pubyear{2021}

\begin{document}
\label{firstpage}
\pagerange{\pageref{firstpage}--\pageref{lastpage}}
\maketitle

\begin{abstract}
%Following the prompt emission of a Gamma-Ray Burst (GRB), a multiwavelength afterglow arises. The afterglow emission has been well modelled by a decelerating relativistic blast wave plunging into the ambient medium and accelerating particles in it, which emit synchrotron radiation. A linear polarization degree of a few percent has been measured in some GRB afterglows in the optical band, and recently a sub-percent detection was also made in the radio. These measurements can offer a wealth of information regarding the afterglow geometry and the critical GRB parameters which we must decipher correctly. In this paper we present a new numerical tool which calculates the time evolution of linear polarization from beamed GRB afterglows. We model the time evolution of the afterglow emission numerically, \del{for different magnetic field structures,} map it on the plane of the sky and integrate over it to obtain the observed polarization signature. We account for relativistic effects as well as how the synchrotron emission and polarization spectrum change in time. We find that for standard afterglow parameters and magnetic field structure, the polarization signature from a relativstic shock is comprised of a double peaked polarization degree and a polarization angle rotation at the polarization degree minimum. 
%Polarimetry is an important tool in probing the magnetic field configuration in GRB afterglows. 
Linear polarization measurements in the optical band show polarization degrees of a few percents at late times. Recently, polarization at sub-percent level was also detected in radio by ALMA, opening the window for multi-wavelength polarimetry and stressing the importance of properly modeling polarization in GRB afterglows across the EM spectrum. 
We introduce a numerical tool that can calculate the polarization from relativistically moving surfaces by discretizing them to small patches of uniform magnetic field, calculating the polarized emission from each cell assuming synchrotron radiation and summing it to obtain the total degree of polarization. 
We apply this tool to afterglow shocks with random magnetic fields confined to the shock plane, considering electron radiative cooling. We analyze the observed polarization curves in several wavelengths above the cooling frequency and below the minimal synchrotron frequency and point to the
characteristic differences between them. We present a method to constrain the jet opening angle and the viewing angle within the context of our model. 
Applying it to GRB 021004 we obtain angles of $\sim10^\circ$ and $\sim8^\circ$ respectively and conclude that a non-negligible component of radial magnetic field is required to explain the $\sim1\%$ polarization level observed $3.5$ days after the burst. 
%\gb{Following the prompt gamma-ray emission of a gamma-ray burst, a multi-wavelength afterglow arises. The emission from the afterglow is well described by a decelerating relativistic forward shock, plowing into the interstellar medium and accelerating particles in it to emit synchrotron radiation. Observations of linear polarization in a few optical afterglows showed a polarization degree of a few percent }
\end{abstract}

\begin{keywords}
gamma-ray burst: general - polarization - methods: numerical
\end{keywords}

%%%%%%%%%%%%%%%%%%%%%%%%%%%%%%%%%%%%%%%%%%%%%%%%%%

%%%%%%%%%%%%%%%%% BODY OF PAPER %%%%%%%%%%%%%%%%%%

\section{Introduction}

A Gamma-Ray Burst (GRB) afterglow (AG) is formed when a relativistic jet plows through the interstellar medium (ISM) and gathers enough material ahead of it to considerably decelerate and dissipate its kinetic energy. 
The interaction of the jet with the ambient medium leads to the formation of two shocks: a forward shock, propagating into the ambient medium, which is responsible for most of the AG emission and a reverse shock, which grows in the ejecta and contributes to the emission at an early stage \citep[e.g.][]{Sari1997Hydro,1999ApJ...513..669K}. 
%The observed emission likely originate from electrons accelerated at the shock front, gyrate around magnetic fields that grew locally in the shock and emitting synchrotron light.
The emission is well described by a broken power-law energy distribution of electrons gyrating around magnetic field lines 
%that grew locally in the \gb{forward} shock 
and emitting synchrotron light \citep[e.g.][]{1993ApJ...418L...5P,1994ApJ...422..248K,1997ApJ...490..772K,1997ApJ...489L..33W,1997ApJ...485L...5W,Sari1998,1998ApJ...499..301M}. 
 %with a fraction $\varepsilon_e$ of the shock energy going to the electrons and a fraction $\varepsilon_B$ that goes into the magnetic field. 
 Though this model is very successful in describing the overall AG emission, detailed properties, such as the configuration of the magnetic field or the acceleration process of the non-thermal particles remain obscured. Such details can help us understand the properties of the ambient medium as well as of the relativistic jets. 
For example, a shock propagating into an unmagnetized medium can grow magnetic field in-situ via local plasma instabilities such as the two-stream Weibel instability \citep[e.g.][]{1999ApJ...526..697M,2005ApJ...618L..75M}. The magnetic field in this case will be mostly tangential to the shock plane with a small coherence length. A radial component may grow downstream of the shock due to plasma motions. On the other hand the medium may contain a non-negligible ordered field component,
as suggested by some models of ISM \citep[e.g.][]{2018A&A...610C...1P} or in the case of a shock propagating in to a magnetized stellar wind \citep[e.g.][]{1993A&A...277..691B}. In this case the shock compressed magnetic field can add a component with a large coherence length to the locally grown random field and alter the field configuration. In a case of a reverse shock, if relics of magnetic field from a magnetically launched jet remain in the upstream plasma, they will be imprinted on the shock and can alter both the particle acceleration process as well as the properties of the observed emission. 

A natural way to probe the properties of magnetic fields in emitting systems is using polarization measurements. Synchrotron radiation from a distribution of particles is linearly polarized in a direction perpendicular to the magnetic field and to the line of sight (LOS). A detection of an overall polarization signature is indicative of a global anisotropy in the magnetic field or in the system geometry. 
A proper modeling of the polarization and how it evolves with time can shed light on the conditions in the emission regions, specifically on the configuration of the magnetic field in the shock and on the particle spectral energy distribution (SED).

So far linear polarization in the AG was observed in the optical band.
The first detections of polarization at a level of $\sim1\%$ in the AGs of GRB 990510 \citep{1999GCN...330....1C,1999ApJ...523L..33W} and GRB 990712 \citep{2000ApJ...544..707R} inspired several analytic works that modeled the polarization assuming a random field configuration on the shock plane 
and synchrotron emitting electrons with a powerlaw SED 
\citep{1999A&A...348L...1C,1999ApJ...524L..43S,Ghisellini1999,1999ApJ...511..852G}. Later observations of GRB AGs with a higher polarization degree, e.g. GRB 020405  \citep{2003ApJ...583L..63B,2003A&A...400L...9C}, motivated models that calculated the polarization from a uniform magnetic field on the shock plane \citep{Granot2003a} and from a random field with a patchy geometrical pattern \citep{2004ApJ...602L..97N}. 
Other detections of polarization showed rotations of the polarization vector over time \citep[e.g.][]{Rol2003A&A,2012MNRAS.426....2W} and changes in the polarization degree measured at different wavelengths \citep{Klose2004A&A}. Lately, polarization at a sub-precentage level was also detected in mm wavelengths using ALMA \citep{Laskar2019ApJ} opening a window for polarization modeling across a wide spectral range. This highlights the importance of modeling both the time evolution as well as the spectral properties of AG polarization.

When modeling the time evolution of the observed polarization, one needs to consider the differences in the light travel times from various regions on the shock \citep[e.g.][]{Sari1997Egg,Granot1999Apj,2008MNRAS.390L..46G}. The effect on the observed polarization and its evolution in time was studied by many authors 
\citep[e.g.][]{Sari1999Pol,Granot2003a,2004MNRAS.354...86R,2019MNRAS.tmp.2582G}.
The spectral properties of the observed image were first calculated analytically by \citet{Sari1999Pol} and by \citet{Granot1999Apj} assuming a single powerlaw SED. 
Lately \citet{Shimoda2020} used these results to obtain the time evolution of the polarization at frequency above and below the synchrotron frequency.    
%Lately, \citet{Shimoda2020} extended these works by accounting for the spectral break at the synchrotron frequencies and how it affects the polarization at frequencies above and below this frequency. 
These studies used analytic descriptions of the system and propagated them in time to obtain the polarization curves. 
We took a different approach of discretizing the emitting zone into individual cells, calculating the time dependent emission and polarization in each cell separately and summing the flux weighted polarization from all cells to obtain the total observed polarization.
A similar approach was taken by \citet{Nava2015a} in calculating the linear and circular polarization in spherical AG shocks with various magnetic field configurations, without accounting for photon travel time effects. 
This method allows us to plot detailed maps of the polarized  images. In addition, it is highly flexible in varying the system properties, introducing asymmetries and adding more physical processes. 
Our method can work with arbitrary magnetic field configurations, viewing angles and particle SED. 
The current version calculates emission from 2D surfaces.
We use it to calculate the polarization accounting for cooling of the emitting particles by using a broken power-law SED and obtain the observed polarization curves in the different spectral regimes. 

We begin by describing the geometrical setup adopted in this work and the different reference frames we use (\S 2). We then discuss how we model the shock emission (\S 3) and the polarization (\S 4). In section 5 we present some indicative results and discuss their implications and differences from other works. Last, we present in \S 6 a method to obtain the observer's viewing angle and the jet opening angle from two observables in the polarization curve. 
 In this work we focus only on the forward shock, and assume a configuration of a random field tangent to the shock plane  and a slow cooling SED. We leave the modeling of other configurations including the emission from reverse shocks to a future work.

\section{Geometrical Setup}
\label{sec:setup}
\begin{figure}

	\includegraphics[width=\columnwidth, trim=0cm 2cm 15cm 0cm]{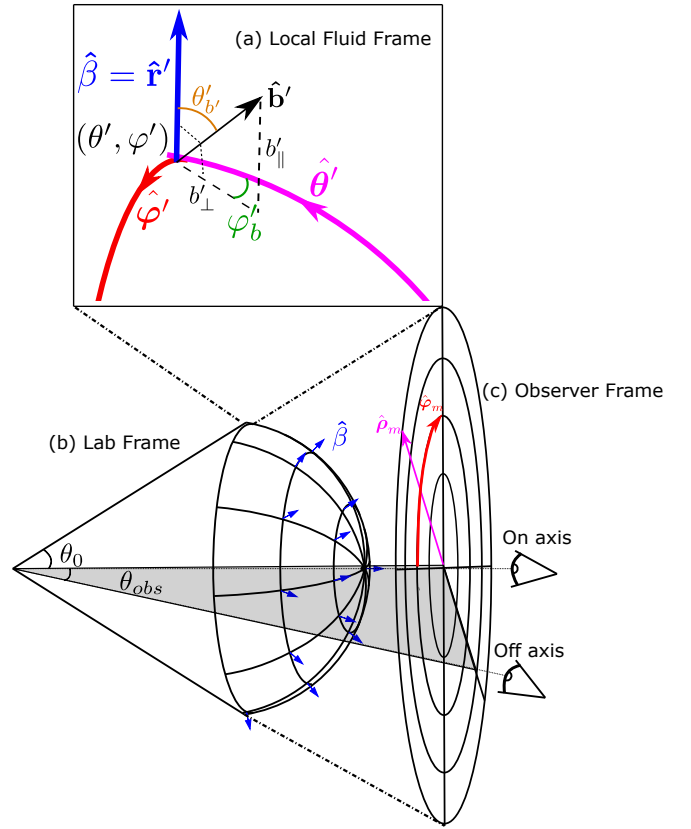}
    \caption{The setup of our system. The local fluid frame is shown in panel (a) with the magnetic field structure denoted by eq. \ref{eq:ShockFrameMF}. Panel (b) shows the AG shock in the lab frame. It has a semi-spherical shape with a half opening angle $\theta_0$. The shock symmetry axis is aligned with the $\unit{z}$ axis. The matter at each point just behind the shock expands radially with a 3-velocity $\vec{\beta}$. Panel (c) shows the observer's map. The image of the AG shock is projected on the map so that the coordinate $\theta$ on the AG shock is mapped into a radial distance on the map as $\rho_m=\sin\theta$ and the coordinate $\varphi$ on the shock is mapped to the coordinate $\varphi_m=\varphi$. The observer is at rest with respect to the lab frame. On-axis observers are aligned with the shock symmetry axis, while an off-axis observer is rotated at an angle $\theta\obs$ from the symmetry axis.}%\obn{we need to fix this plot so that the diameter of the map equals the diameter of the shock.}}
    \label{fig:SystemSetup}
\end{figure}
Our system consists of a spherical-cap shaped blast-wave with a half opening angle $\theta_0$, propagating in a medium and driving a shock ahead of it. We assume that the observed emission comes from the fluid just behind the shock moving at a Lorentz factor $\Gamma=\Gamma_{\rm sh}/\sqrt{2}$, where $\Gamma_{\rm sh}$ is the Lorentz factor of the shock\footnote{In practice the emission comes from a layer of thickness $\sim R/\Gamma^2$ having a spread of Lorentz factors \citep[see e.g.][]{1993MNRAS.263..861P}, however since our model is 2D we assume that all the emission comes from an infinitely thin surface that coincides with the shock surface.}. 
The symmetry axis is aligned with the $\unit{z}$ axis, while the observer can be aligned with the jet axis (on-axis observer) or misaligned by an angle $\theta\obs$ from it (off-axis observer).
%The observer is located on the $z$ axis, and the jet symmetry axis is either aligned with the observer (on-axis jet) or miss aligned by an angle $\theta_{\rm obs}$ (off-axis jet). 
We consider three inertial frames: i) The local fluid frame at the shock immediate downstream, located at the same radius as the shock. ii) The lab frame, in which the ambient medium is at rest. iii) The observer frame. We neglect cosmological expansion, thus the observer is at rest with respect to the lab frame.
We use tagged values for quantities in the local fluid frame, untagged values for lab frame quantities and sub index "${\rm obs}$" to mark quantities in the observer frame. 

\subsection{Local fluid frame geometry}
%We conduct most of our calculations in the shock frame. Quantities in the lab (identical to the ISM) frame are marked with sub index "${\rm ISM}$" and values in the observer frame with sub index "${\rm obs}$". We neglect cosmological expansion, thus the observer is at rest with respect to the lab frame.} \obn{change markings according to this.
The magnetic field in the local frame, $\vect{b'}$, is defined with the following two angles:
%The direction of the shock rest frame magnetic field $\bf{\hat{b}}'$ in is described with the following two angles: %\obn{small letters are commonly used for local frames and capital letters for lab frame} 
\begin{enumerate}
	\item $\theta'_{\rm b}\in[0,\frac{\pi}{2}]$ determines the angle of $\vect{b'}$ from the local radial direction. So that ${b}'_\perp\equiv b'\sin\theta'_{\rm b}$ is the magnetic field component on the shock surface.
	\item $\varphi'_{\rm b} \in[0,2\pi]$ measures the
	orientation of $\vect{b}'_\perp$ in the azimuthal direction ($\unit{\varphi}'$) from the local $\unit{\theta}'$ direction, namely $\cos{\varphi'_{\rm b}}=\unit{b}'_\perp\cdot\unit{\theta}'$. The two angles are illustrated in fig. \ref{fig:SystemSetup}a. 
\end{enumerate}
With these definitions, a unit vector of the magnetic field in the local frame is defined as
\begin{equation}
\unit{b}'=\cos\theta'_{\rm b}\bf{\hat{r}}'+\sin\theta'_{\rm b}\cos\left(\varphi'+\varphi'_{\rm b}\right)\boldsymbol{\hat{\theta}}'+\sin\theta'_{\rm b}\sin\left(\varphi'+\varphi'_{\rm b}\right)\boldsymbol{\hat{\varphi}}'
\label{eq:ShockFrameMF}
\end{equation}
%\ob{are you certain that it shouldn't be $\sin\theta'_{b}\cos\left(\phi'_b\right)\boldsymbol{\hat{\theta}}$ and $\sin\theta'_{b}\sin\left(\phi'_b\right)\boldsymbol{\hat{\phi}}$?}

The choice of $\theta'_{\rm b}$ and $\varphi'_{\rm b}$ allows us to define arbitrary configurations for the magnetic field. In this work we consider a random magnetic field on the plane of the shock, thus we use $\theta'_{\rm b}=\frac{\pi}{2}$ and and randomize $\varphi'_{\rm b}$ at each cell.

\subsection{Observer coordinate system and alignment}
The observer's map is modeled as a projection of the spherical shock on a plane perpendicular to the jet axis so that the projected angular size is preserved\footnote{This is not the image an observer sees, but a convenient way to map each angle on the shock to a unique location on the observer's plane to assess and illustrate its contribution to the total polarization} (see fig. \ref{fig:SystemSetup}b,c for illustration).
The projected image is a circle centered around the jet axis, with 2D polar coordinates: $\rho\uum=\sin\theta$, $\varphi\uum=\varphi$ with a differential surface area 
\begin{equation}
dS_{\rm m}=\rho\uum d\rho\uum d\varphi\uum=\cos\theta\sin\theta d\theta d\phi=d\Omega_\perp.
\end{equation}
%We map the lab frame shock coordinate $\theta$ to a radial coordinate on the observer plane $\rho_m\equiv\theta$. The shock coordinate $\varphi$ is mapped to the azimuthal map coordinate $\varphi_m$. 
For an on-axis observer, the LOS points in the $\unit{z}$ direction. For any coordinate ($\rho\uum,\varphi\uum$) on the observer's map, the unit vector at the corresponding location on the shock pointing at the observer is
\begin{equation}
\unit{n}\obs=\cos\theta\unit{r} - \sin\theta\unit{\theta}.
\label{n_obs_onAxis}
\end{equation}
To change the observer's viewing angle, we apply a rotation matrix, $R(\Psi,\unit\kappa)$, where $\unit\kappa$ is the axis of rotation, $\Psi$ is the rotation angle and the rotation is done according to the right hand rule with respect to $\unit\kappa$. In this work, we assume that all rotations of the observer are done about the $\unit{y}$ axis. Therefore, the unit vector of an off-axis observer is 
$
\unit{n}\obs=R(\theta\obs,\unit{y})\unit{z},
$
where $\theta\obs\equiv q\theta_0$ and $0\leq q\leq1$. The explicit expression for $\unit{n}\obs$ on the shock surface is therefore
\begin{equation}
\begin{aligned}
	\unit{n}\obs&=(\sin\theta\cos\varphi\sin\theta\obs+\cos\theta\cos\theta\obs)\unit{r}\\
	&+(\cos\theta\cos\varphi\sin\theta\obs-\sin\theta\cos\theta\obs)\unit{\theta}\\
	&-\sin\varphi\sin\theta\obs\unit{\varphi}.
	\label{eq:n_obs_offAxis}
\end{aligned}
\end{equation}
It can be seen that for $\theta\obs=0$ the expression is equal to eq. \ref{n_obs_onAxis}. 
%An off-axis observer deflected by  and angle $\theta_{\rm obs}=q \theta_0$, will show on the map at location $\rho_m=q$.
The mapping as well as the location of the observer are illustrated in fig. \ref{fig:SystemSetup} on panels b and c.
%See panels (b) and (c) in Fig. \ref{fig:SystemSetup} for an illustration of the mapping.

\subsection{Transformations of the observer LOS to the local fluid frame}
\label{sec:transformation}
To calculate the direction to the LOS in the local frame, we rotate $\unit{n}\obs$ by an angle 
\begin{equation}
%\xi=\cos^{-1}\left(\frac{\unit{\beta}\cdot\unit{n}\obs-\beta}{1-\vect{\beta}\cdot\unit{n}\obs}\right)-\cos^{-1}\left(\unit{\beta}\cdot\unit{n}\obs\right),
\xi=\cos^{-1}\left(\frac{\mu\obs-\beta}{1-\beta\mu\obs}\right)-\cos^{-1}\left(\mu\obs\right),
\label{eq:xi_rotation}
\end{equation}
about the axis 
\begin{equation}
\unit{\xi}=\unit\beta\times\unit{n}\obs,
\end{equation}
where $\mu\obs=\unit{\beta}\cdot\unit{n}\obs$ depends on both $\theta$ and $\varphi$ in the general case of an off-axis observer.
The rotation is done by applying the rotation matrix, $R(\xi,\unit{\xi})$, obtaining  
\begin{equation}
\unit{n}'\obs=R(\xi,\unit{\xi})R(\theta\obs,\unit{y})\unit{z}.
\label{n_obs_transform}
\end{equation}
This expression is a generalization of the classic aberration of light for an arbitrary viewing angle. In the case of an on-axis observer ($\theta\obs=0$),
 we get that $\mu\obs=\cos\theta$ and the rotational angle is just
 \begin{equation}
%\xi=\cos^{-1}\left(\frac{\cos\theta-\beta}{1-\beta\cos\theta}\right)-\theta=
\xi=\theta'-\theta.
\end{equation}
The rotation in this case is done along meridian lines about the $\unit{\xi}=-\unit\varphi$ axis. The transformation from the shock frame back to the lab frame is done by rotating the LOS in the opposite direction, ($\unit{n}'\obs\times\unit{\beta}$) and expressing $\xi$ in terms of $\vect{\beta}\cdot\unit{n}'\obs$

\section{Emission}
\label{sec:Emission} 

\begin{figure}	
	\includegraphics[width=\columnwidth, trim=3cm 9.5cm 3cm 9.5cm]{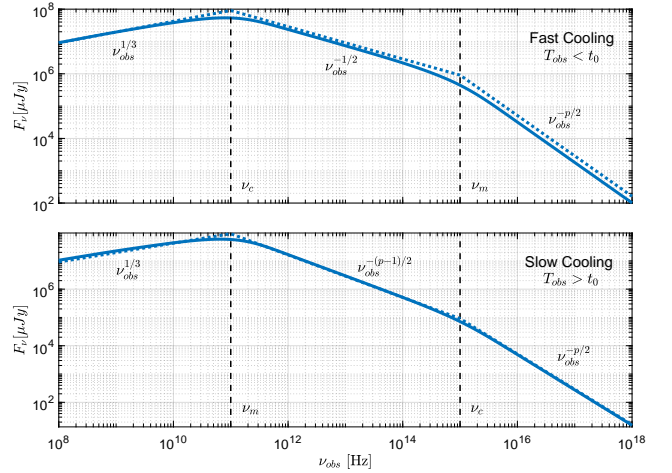}
	\caption{Spectral flux from a powerlaw distribution of electrons with a powerlaw index of $p=2.5$ in two scenarios: 
	i) Top panel: Fast cooling scenario, with $\nu_c<\nu_m<\nu_{\rm max}$.
	ii) Bottom panel: Slow cooling scenario, with $\nu_m<\nu_c<\nu_{\rm max}$. 
	%\gb{The spectrum is calculated for the slow cooling AG model, where the critical frequencies were set artificially.} 
	The locations of the critical frequencies $\nu_m,\nu_c$ are marked with black vertical dashed lines. The solid blue lines show our smooth model while the dotted blue lines are the piecewise synchrotron spectrum used by \citet{Sari1998}. The analytic spectrum is multiplied by 8 to match the numerical one. The origin of the factor 8 difference is discussed in appendix B.}
	\label{fig:EmissionSpectrum}
\end{figure}

The emission is calculated in the local fluid frame and boosted to the observer frame. 
%We follow the calculation of \citet{Sari1997} who considered 
Following the calculation of \citet{Sari1998} we consider a relativistic forward shock decelerating self-similarly in a medium with a uniform density (the ISM), according to the Blandford \& McKee solution \citep{1976PhFl...19.1130B}. 
Electrons are accelerated on the shock front to a power-law  distribution in energy and emit synchrotron radiation in the presence of magnetic fields generated on the shock. The magnetic field in the local frame is parameterized as:
\begin{equation}
\vect{b}'=(32\pi  \varepsilon_{_b} n m_p)^{\frac{1}{2}}\Gamma c \unit{b}',
\label{eq:MF}
\end{equation}
 where $m_p$, $n$ are the proton mass and number density in the ambient medium respectively and $\varepsilon_{_b}$ is the fraction of shock energy that goes to the magnetic field. 
 %and $\Gamma$ is the Lorentz factor of the downstream fluid just behind the shock. It relates to the shock Lorentz factor via $\Gamma={\Gamma_{\rm sh}}.{\sqrt{2}}$.
The synchrotron power per unit frequency of a single electron is given by \citep{Rybicki1979}
\begin{equation}
\dot{E}'_{e,\nu'}(\gamma)=\frac{\sqrt{3}e^3b'\sin\alpha'}{m_e c^2}\tilde{F}\left(\frac{\nu'}{\nu'_{\rm s}(\gamma)}\right),
\label{eq:P_numax}
\end{equation} 
where $\tilde{F}\left(x\right)=x\int^{\infty}_x K_{\frac{5}{3}}\left(\xi\right)d\xi$, is the integrated modified Bessel function of order $\frac{5}{3}$, 
\begin{equation}
\nu_{\rm s}'\left(\gamma\right)=\frac{3\gamma^2eb'\sin\alpha'}{4\pi m_e c}
\label{eq:SynchFreq}
\end{equation}
is the synchrotron frequency expressed with $\gamma$, the electron Lorentz factor in the shock frame and $\alpha'=\cos^{-1}({\unit{b}'}\cdot\unit{n}'\obs)$ is the pitch angle of the electrons radiating into the LOS. 
Note that although $\gamma$ is defined in the local frame we leave it untagged. 
For the electron distribution we use the standard fast-cooling and slow-cooling distributions \citep{Sari1998}.
The electron number density is assumed to be uniform in the flow. It has
%We take a uniform electron number density in the flow with 
a broken powerlaw distribution in $\gamma$ between a minimal Lorentz factor $\gamma_m\propto\Gamma$ and a maximal value $\gamma_{\rm max}$ with a break at $\gamma_c$, above which electrons cool over dynamical timescales and the particle distribution steepens. The fast-cooling distribution is relevant for early times. It is characterized by $\gamma_c<\gamma_m<\gamma_{\rm max}$ 
and has an electron number density distribution of 
\begin{equation}
n'_\gamma\propto \begin{cases}
\gamma^{-2}&\gamma_c<\gamma<\gamma_m\\
\gamma_m^{p-1}\gamma^{-p-1}&\gamma_m<\gamma<\gamma_{\rm max}\\
\end{cases}.
\label{eq:ParticleSpectrumSFast}
\end{equation}
%ii) Slow cooling with 
%$\gamma_m<\gamma_c<\gamma_{\mbox{max}}$, associated with later times in the AG life and is characterized by electron number density distribution of
Slow-cooling occurs at later times, is characterized by $\gamma_m<\gamma_c<\gamma_{\rm max}$ and has an electron number density distribution of 
\begin{equation}
n'_\gamma\propto \begin{cases}
\gamma^{-p}&\gamma_m<\gamma<\gamma_c\\
\gamma_c\gamma^{-p-1}&\gamma_c<\gamma<\gamma_{\mbox{max}}\\
\end{cases}.
\label{eq:ParticleSpectrumSlow}
\end{equation}
The total emitted power per unit frequency is calculated by integrating the power per unit frequency of a single electron (eq. \ref{eq:P_numax}) over the entire electron population in the shock,
\begin{equation}
\dot{E}'_{\nu'}=\int \dot{E}'_{e,\nu'}(\gamma)N_\gamma d\gamma,
\label{eq:SpectrumShape}
\end{equation} 
where $N_\gamma=C n'_\gamma$ is the total number of electrons per unit $\gamma$. The normalization coefficient $C$ can be obtained by noting that the total number of radiating electrons is equal to the total number of electrons swept up by the shock, i.e. $C\int n'_\gamma d\gamma=\frac{2\pi(1-\cos\theta_0)}{3} R^3 n$.
%, where $n'_\gamma$ is the number density of the fast cooling or slow cooling electrons, depending on the shock conditions. 
Assuming an isotropic distribution of magnetic field on the shock plane and of electron velocity we get that the 
%Since the emission is isotropic in the fluid frame, the 
specific intensity is 
$I'_{\nu'}=\dot{E}'_{\nu'}/2\pi(1-\cos\theta_0) R^2$.
Transforming to the observer frame and noting that $I_{\nu}=D^3I'_{\nu/D}$ we get:
\begin{equation}
I_{\nu}=D^3\frac{Rn}{3}\frac{\int \dot{E}'_{e,\nu/D}(\gamma)n'_\gamma d\gamma}{\int n'_\gamma d\gamma}\propto D^{3-\kappa}R(b'\sin\alpha')^{1-\kappa}\gamma_m^{-2\kappa},
\label{eq:ObsIntensity}
\end{equation}
where $D=\left[\Gamma(1-\beta\mu\obs)\right]^{-1}$ is the Doppler factor of the shocked fluid and $\kappa$ is the spectral slope at frequency $\nu/D$.
To integrate the spectral energy density in the numerator we use analytic approximations that hold both far and close to the critical frequencies $\nu'_{m,c}\equiv \nu'_s(\gamma_{m,c})$, 
as illustrated in fig. \ref{fig:EmissionSpectrum}. This is opposed to the piecewise solution used in many analytic models \citep[e.g.][]{Sari1998}.
%The integration of the spectral energy density at the numerator is preformed numerically for $\nu'$ close to the critical frequencies $\nu'_{m,c}\equiv \nu'_s(\gamma_{m,c})$ and analytically elsewhere\footnote{See \citet{Rybicki1979} eq. 6.36 for the analytic approximation.}. \gb{I think I commented this before, I approximate $\tilde{F}(x)$ everywhere} This way we obtain a smooth spectral shape around the critical frequencies illustrated in fig. \ref{fig:EmissionSpectrum}, as opposed to the piecewise function used in many analytic models \citep[e.g.][]{Sari1998}. 
The difference between the two solutions becomes important when calculating the polarization spectrum as we show in sec. \ref{ssec:PolSpect}.

%We map the radiation from each angular element on the emitting surface to the corresponding angle on the observer's plane. In order to evaluate the observed image at each cell on the map
The observer's map in our model is an angular projection of the emitting surface on the sky. In order to evaluate the observed image in each map cell we need to calculate the specific flux per unit angle  
%We map the observed radiation from each angular element on the emitting surface to the corresponding angle on the observer's map. Therefore we are interested in the specific flux per unit angle 
on the plane of the sky (surface brightness) that falls on each cell, defined as \citep[e.g.][]{Sari1997Egg}
\begin{equation}
\frac{dF_\nu(\rho\uum,\varphi\uum)}{dA}=I_\nu\left(\frac{R}{d_{_{\rm L}}}\right)^2\mu\obs\frac{d_{_{\rm L}}^2}{R_\perp}\frac{d\mu\obs}{dR_\perp},
\label{eq:dFdS}
\end{equation}
where $R_\perp=R\sqrt{1-\mu\obs^2}$ is the perpendicular distance of the emitting element from the LOS and
$dA=\frac{R_\perp dR_\perp d\varphi}{d_{_{\rm L}}^2}$ is a differential solid angle on the plane of the sky at the emission point. 
The LHS of eq. \ref{eq:dFdS} is evaluated at the map coordinate $(\rho\uum,\varphi\uum)$, while the RHS is calculated at the corresponding shock coordinates $(R,\theta,\varphi)$.
The value of $\frac{d\mu\obs}{dR_\perp}$ is determined by the shape of the emitting surface (see below). 
The total flux on the map is obtained by integrating the differential flux 
\begin{equation}
dF_\nu(\rho\uum,\varphi\uum)=\frac{dF_\nu}{dA}dA= I_\nu\left(\frac{R}{d_{_{\rm L}}}\right)^2\mu\obs d\Omega\simeq I_\nu\left(\frac{R}{d_{_{\rm L}}}\right)^2dS_{\rm m},
\label{eq:dF_nu}
\end{equation}
over all cells in the map. 
Note that when the observer is located at an angle $\theta\obs$ from the jet axis,
the map in our model is slightly misaligned with the LOS and the flux needs to be multiplied by an additional $\cos\theta\obs$. However since in this work $\theta\obs\leq\theta_0\ll1$ the effect is negligible and we ignore it. 

\subsection{Photon arrival times}
\label{ssec:flux_calc}
Due to the relativistic motion, two photons emitted 
from the same point on the shock at a time difference $\delta t$ in the lab frame reach the observer in a time interval 
\begin{equation}
\delta T\obs = \delta t(1 - \beta_{{\rm sh}}\mu\obs) %\vect{\beta}\cdot\unit{n}\obs).
\label{eq:dt_obs}
\end{equation}
The difference between the emission time interval and the observed one leads to a mixing of photons emitted over a range of lab times and arrive to the observer simultaneously. %\gb{The difference between the lab time interval and the observed one causes photons, emitted over a range of lab times and radii, to arrive simultaneously at the observer frame.}
%This modifies shape of the observed surface from the spherical shock surface 
This modifies the observed shape of the emitting surface from spherical to an egg-like shape denoted as the equal arrival time surface (EATS).
%\footnote{In practice photons are emitted from a layer behind the shock with a thickness $\sim R/\Gamma^2$, resulting in a volume from which photons are arriving simultaneously to the observer \citep[see e.g.][]{Granot1999Apj,2008MNRAS.390L..46G}. However since in our model we assume an infinitesimally thin emitting surface, the region of equal photon arrival time in our case is a surface. \gb{This is the exact same as footnote 1...}} (EATS). 

\begin{figure}
    \includegraphics[width=\columnwidth, trim=1.8cm 2cm 2cm 2.4cm]{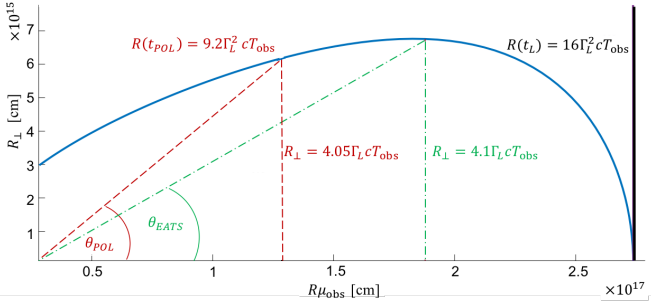}
\caption{The shape of the EATS (blue), obtained from eq. \ref{eq:EAT_surface}, plotted 
on a plane perpendicular to the LOS at time $T\obs=0.0602$ days. 
Radiation is emitted from each point on the EATS at a different lab time and reaches the observer at a time $T\obs$.
The black vertical line is the shock surface at the lab time 
associated with $T\obs$ on the LOS and 
$R(t_{_{\rm L}})=16\Gamma_{_{\rm L}}^2 cT\obs$ is it's radius.
The EATS maximum is located at a height $R_\perp\simeq4.1\Gamma_{_{\rm L}}cT\obs$ (green dash-dotted line), and forms an opening angle of $\theta\uuEATS\simeq0.4/\Gamma_{_{\rm L}}$ from the LOS. 
It divides the EATS to a front part and a back part located on the right and on the left of this point respectively.
A second important angle, $\theta\uuPOL\simeq0.45/\Gamma_{_{\rm L}}$ (crimson dashed line), is the angle on the EATS for which $\theta\uuPOL=1/\Gamma(\theta\uuPOL)$. On this angle the observed polarization is completely radial. 
}
\label{fig:shape_EATS}
\end{figure}

We define $t$ as the time measured in the lab frame from the onset of the GRB and $T\obs$ as the time in the observer frame that passed from the arrival of the first photon emitted at $t=0$. To calculate the shape of the EATS we integrate eq. \ref{eq:dt_obs} and find the relation between $T\obs$ and $t$ for a case of a decelerating shock. 
We assume that the shock decelerates adiabatically, where the shock radius scales with it's Lorentz factor  as $R\propto\Gamma_{\rm sh}^{-3/2}$ \citep[e.g.][]{Sari1997Hydro}. %Approximating $\beta_{\rm sh}\simeq1-\frac{1}{2\Gamma^2}$, $\mu\obs\simeq1-$, 
Expanding $\beta_{\rm sh}$ and $\mu\obs$ to first order, taking $R=ct$ and expressing the solution in terms of $\Gamma=\Gamma_{\rm sh}/\sqrt{2}$, the shocked fluid Lorentz factor, the integration of eq. \ref{eq:dt_obs} gives  \citep{Sari1997Egg,Granot1999Apj}:
\begin{equation}
%T\obs=t\left(1-\mu\obs+\frac{\mu\obs}{16\Gamma^2}\right).
T\obs=t\left(1-\mu\obs+\frac{1}{16\Gamma^2}\right).
\label{eq:t_obs_mod}
\end{equation}
and the associated radius is
\begin{equation}
%R(T\obs,\mu\obs)=\frac{cT\obs}{1-\mu\obs +\frac{\mu\obs}{16\Gamma^2}}.
R(T\obs,\mu\obs)=\frac{cT\obs}{1-\mu\obs +\frac{1}{16\Gamma^2}}.
\label{eq:EAT_surface}
\end{equation}
%where $\mu\equiv\unit{\beta}\cdot\unit{n}\obs$. 
The shape of the EATS is obtained by fixing $T\obs$ and calculating $R(T\obs,\mu\obs)$.
To connect the shock radius with the fluid Lorentz factor we calculate their value on the LOS, using the velocity profiles obtained from the Blandford-Mckee solution and assuming that the total energy $E$ in the flow is conserved \citep{Sari1998}:
\begin{equation}
R_{_{\rm L}}(T\obs)=\left(\frac{17ET\obs}{\pi m_p nc}\right)^{\frac{1}{4}},
\label{eq:R_t}
\end{equation}
\begin{equation}
\Gamma_{_{\rm L}}(T\obs)=\frac{1}{4}\left(\frac{17E}{\pi nm_p c^5 T\obs^3}\right)^{\frac{1}{8}}.
\label{eq:Gamma_t}
\end{equation}
With these quantities $R$ and $\Gamma$ maintain
\begin{equation}
\Gamma=\Gamma_{_{\rm L}}\left(\frac{R_{_{\rm L}}}{R}\right)^{\frac{3}{2}}.
\label{eq:RG}
\end{equation}
An analytic solution to $R(T\obs,\mu\obs)$ was given by \citet{Sari1997Egg,Granot1999Apj} for an on-axis observer.
%and approximating $\mu\obs\simeq1-\frac{\theta^2}{2}$. \gb{I would remove the approximaton bit, as I see you removed the part about where we don't approximate $\mu$ in $R(\mu,t)$} 
Figure \ref{fig:shape_EATS} shows the shape of the EATS on the $r-\theta$ plane calculated from eqn. \ref{eq:R_t}-\ref{eq:RG} at time $T\obs=0.0602$ days (see model parameters in sec. \ref{sec:results}). 
It is consistent with the shapes obtained in previous works. 
%To illustrates the difference between the shock surface and the EATS we added plots of the shock surface at different lab times.
%ith The green dashed line marks the analytic solution of \citet{Sari1997Egg}. 
%We add several shock surfaces at different observed times on the LOS to demonstrate the difference between emitting surfaces and the EATS.

To calculate the observed surface brightness we need to calculate the quantity $\frac{1}{R_\perp}\frac{d\mu\obs}{dR_\perp}$ on the EATS. Using the above relations we obtain (see appendix A for for the derivation)
%\citep[see][for a step by step derivation of this quantity]{Sari1997Egg}
\begin{equation}
\frac{1}{R_\perp}\frac{d\mu\obs}{dR_\perp}=\frac{1}{R^2}\left(\frac{1+3\left(\frac{R}{R_{_{\rm L}}}\right)^4}{1-5\left(\frac{R}{R_{_{\rm L}}}\right)^4}\right).
\label{eq:dmu_dS}
\end{equation}
Substituting that in eq. \ref{eq:dFdS} and calculating the corresponding $R(T\obs,\mu\obs)$ and $\Gamma(T\obs,\mu\obs)$ according to eqn. \ref{eq:R_t}-\ref{eq:RG}, we get the observed surface brightness at each cell on the map,
\begin{equation}
\frac{dF_\nu(\rho\uum,\varphi\uum)}{dA}=I_\nu\left(\frac{1+3\left(\frac{R}{R_{_{\rm L}}}\right)^4}{1-5\left(\frac{R}{R_{_{\rm L}}}\right)^4}\right)\mu\obs,
\label{eq:dFdS_EATS}
\end{equation}
where $I_\nu$ is given in eq. \ref{eq:ObsIntensity}. In appendix \ref{app:ComparisonToAnalytic} we show a comparison of $\frac{dF_\nu(\rho\uum,\varphi\uum)}{dA}$ to the analytic expression obtained in \citep{Sari1997Egg}.%\obn{not existing at the moment}
%For each observed time and a cell on the observer map we calculate the corresponding $R(t\obs,\mu\obs)$ and $\Gamma(t\obs,\mu\obs)$ according to eqn. \ref{eq:R_t}-\ref{eq:Gamma_t}. Although the EATS shape is non-spherical, the emission at each location on the EATS originates from a spherical surface, thus the intensity and fluxes are calculated in a similar manner as in eqn. \ref{eq:ObsIntensity}-\ref{eq:dF_nu}.
%We then sum the differential fluxes at all cells on the map to obtain the total observed flux. 

\section{Polarization}
\label{sec:PolCalc}

%\subsection{Polarization spectra from a uniform field}
\subsection{Polarization spectrum in a uniform field}
\label{ssec:PolSpect}
\begin{figure}

	\includegraphics[width=\columnwidth, trim=4cm 9.5cm 3cm 9.5cm]{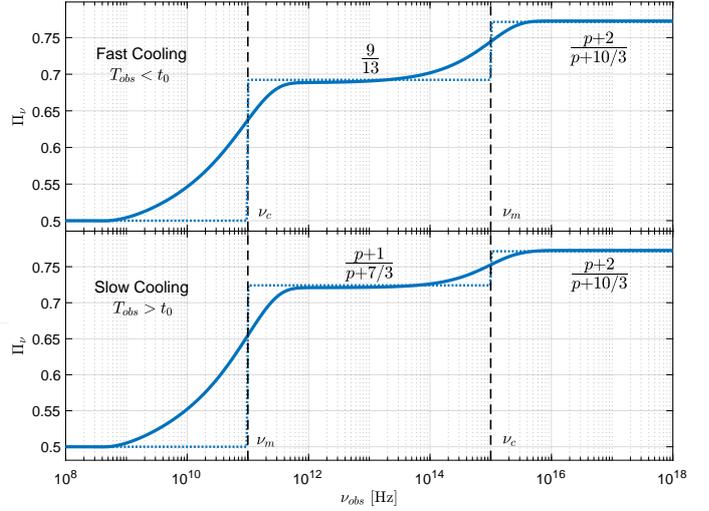}
	\caption{Polarization spectrum for a powerlaw distribution of electrons with a powerlaw index of $p=2.5$ in two scenarios. 
	i) Top panel: Fast cooling scenario, with $\nu_c<\nu_m<\nu_{\rm max}$.
	ii) Bottom panel: Slow cooling scenario, with $\nu_m<\nu_c<\nu_{\rm max}$. %\gb{The polarization spectrum is calculated for the slow cooling AG model, where the critical frequencies were set artificially.} 
	The locations of the critical frequencies $\nu_m,\nu_c$ are marked with black vertical dashed lines. Solid blue lines depicts our semi-numeric solution, while the analytic step-function approximation is shown in dotted blue lines.}
	\label{fig:PolSpectrum}
\end{figure}

While synchrotron radiation from a single particle is elliptically polarized,
emission from a group of particles with a smooth pitch angle distribution is linearly polarized.  
For a group of particles with a distribution of Lorentz factors gyrating around a uniform magnetic field, the polarization degree can be expressed as \citep{Rybicki1979}.
%\gb{This is the formula for integration over the particle population. Thus, this represents the polarization of a particle population $n_{\gamma}$. The PD of a single energy $\gamma$ is $\Pi(\nu)=\frac{G(x)}{F(x)}$ where $x=\frac{\nu}{\nu_{\text{synch}}}$ (eq. 6.37 \citep{Rybicki1979})}, gyrating around a uniform magnetic field, the polarization degree can be expressed as \citep{Rybicki1979}:
\begin{equation}
\Pi_\nu=\frac{\int \tilde{G}\left(\frac{\nu}{\nu_{\rm s}}\right) dn_\gamma} {\int \tilde{F}\left(\frac{\nu}{\nu_{\rm s}}\right) dn_\gamma},
\label{eq:PolSpectrum}
\end{equation}
%\gb{I would write this equation as $\Pi=\frac{\int \tilde{G}\left(\frac{\nu}{\nu_{sync}}\right)N_\gamma d\gamma}{\int \tilde{F}\left(\frac{\nu}{\nu_{sync}}\right)N_\gamma d\gamma}$ to be consistent with eq. \ref{eq:SpectrumShape}, if you want to talk about polarization from a particle population}
where $\tilde{G}\left(x\right)=xK_{\frac{2}{3}}\left(x\right)$. In case of a powerlaw distribution of particles, $n(\gamma)\propto \gamma^{-p}$, the polarization degree can be approximated as $\Pi=\frac{p+1}{p+7/3}$ far from the distribution edges \citep{Rybicki1979}. Using this approximation, \cite{Granot2003} fitted a polarization spectrum to the piecewise fast and slow cooling synchrotron spectra, obtaining step function solutions with jumps occurring at the various critical frequencies (fig. \ref{fig:PolSpectrum}, dashed lines). We refined this calculation by evaluating $\Pi$ over the smooth emission function we obtained in Section \ref{sec:Emission}, using the same method for integrating the modified Bessel functions. 
Fig. \ref{fig:PolSpectrum} shows a comparison of our method to the analytic approximation of \citet{Granot2003} for cases of fast cooling (upper panel) and slow cooling (low panel) synchrotron spectra. Our solution (solid lines) converges to the analytic model (dotted lines) far from the critical frequencies and changes gradually over a range of $\sim2$ orders of magnitudes in frequencies close to them. A substantial difference from the analytic solution is seen at low frequencies, below $\nu_m$($\nu_c$) in the slow (fast) cooling spectrum. The difference can be important for measurements in the optical band at early times or at microwave-radio band at late times as we show below.

A change in the polarization degree can be seen if the polarization is measured instantaneously in several frequencies above and below a critical frequency (fig. \ref{fig:PolSpectrum}). Alternatively, it can also be seen when measuring the polarization in a single frequency over time, if during this time the observed frequency is crossed by a critical frequency. As the forward shock decelerates, the critical frequencies shift to lower values and may cross the observed frequency \citep{Sari1998}. We mark by $t_m (t_c)$ the time when $\nu_m (\nu_c)$ drops below the observed frequency. A third important time, marked by $t_0$, occurs when $\nu_c$ drops below $\nu_m$ and the synchrotron spectrum shifts from fast cooling to slow cooling. Beyond this time the spectral slope for frequencies between $\nu_m$ and $\nu_c$ changes (see fig. \ref{fig:EmissionSpectrum}) and the polarization degree changes accordingly.

\subsection{The polarization vector in a general afterglow field}
The polarization degree of light emitted by a single element with a uniform magnetic field is set by the spectral energy distribution of the emitting particles (fig. \ref{fig:PolSpectrum}). In the general case, both the polarization vector and the polarization degree may vary between different emitting regions, altering the total polarization signature.
%\del{The total polarization is the sum of the polarized light from all radiating regions in the shock} \gb{The total polarization is the polarized fraction of the total emission}. 
%In a uniform magnetic field the polarization vector all regions emit polarized light in the same direction and the polarization degree is set only by the spectral energy distribution of the emitting particles (fig. \ref{fig:PolSpectrum}). In a non-uniform field the polarization vector varies according to the local field direction, altering the total polarization signature. 
To account for this effect we need to properly sum the contribution of polarized light from all regions.
%We follow the method of \citet{Nava2015a} and generalize their calculation for arbitrary configurations of magnetic fields. 
We calculate the direction of polarization in the local frame, transform it to the observer frame and sum the flux weighted contributions from all regions to obtain the total polarization. Our method follows the calculation of \citet{Nava2015a} and generalizes it for arbitrary  magnetic field configurations. 

The direction of the polarization vector in each cell in the local frame is defined as 
\begin{equation}
\unit{P}_0'=\unit{n}'\obs\times\unit{b}',
\label{eq:nXB}
\end{equation}
where $\unit{b}'$ is the unit vector of the magnetic field in a cell and $\unit{n}'\obs$ is the direction to the observer in the local frame.
%Let us first demonstrate the calculation for an on-axis observer where we can get an analytic expression to the local polarization angle in the observer frame and then show the general calculation. 
For an on-axis observer, we can get an analytic expression to the local polarization angle in the observer frame.
Taking the direction to the LOS in the local frame:
 \begin{equation}
\unit{n}'\obs=\cos\theta'\unit{r}'-\sin\theta'\unit{\theta}'
\label{nobs_shock_LOS}
 \end{equation} 
 together with the local frame magnetic field
\begin{equation}
\unit{b}'=\cos\theta'_{\rm b}\unit{r}'+ \sin\theta'_{\rm b} \cos\left(\varphi'+\varphi'_{\rm b}\right)\boldsymbol{\hat{\theta}}'+ \sin\theta'_{\rm b}\sin\left(\varphi'+\varphi'_{\rm b}\right)\boldsymbol{\hat{\varphi}}'
\label{eq:ShockFrameMF1}
\end{equation}
we get the direction of the polarization vector in the local  frame,
\begin{equation}
\begin{split}
\unit{P}_0'= &-\left[\sin\theta'_{\rm b}\sin\left(\varphi'+\varphi'_{\rm b}\right)\right]\unit{\psi}'\\
&+\left[\cos\theta'\sin\theta'_{\rm b}\cos\left(\varphi'+\varphi'_{\rm b}\right)+\sin\theta'\cos\theta'_{\rm b}\right]\unit{\varphi}',
\end{split}
\end{equation}
where $\unit{\psi}'=\sin\theta'\unit{r}'+\cos\theta'\unit{\theta}'$ is a unit vector on the $\vect{r}'-\vect{\theta}'$ plane perpendicular to $\unit{n}'\obs$ and to $\unit{\varphi}'$. The polarization vector remains perpendicular to the LOS and to the magnetic field at any reference frame.
Any rotation applied on $\unit{n}'\obs$ (and $\unit{b}'$) will rotate $\unit{P}_0'$ in the same way.
%Therefore any rotation applied on $\unit{n}'\obs$ is applied on $\unit{P}_0'$ as well.
%The polarization points in the direction of the wave electric field, which is in the direction of $\unit{n}\obs\times\unit{b}$ in any reference frame. 
The transformation of $\unit{n}'\obs$ to the observer frame is obtained by applying 
%From the discussion in sec. \ref{sec:transformation}, we can calculate the transformation of $\unit{n}'\obs$ to the observer frame by applying 
the rotational matrix $R(\xi,\unit{\varphi}')$, which rotates $\unit{n}'\obs$ by an angle $\xi=\theta'-\theta$ about the $\unit{\varphi}'$ axis (see sec. \ref{sec:transformation}).
The same rotation applied on $\unit{P}_0'$ rotates only the $\unit{\psi}'$ component by the same angle. Since $\unit{\psi}'\perp(\unit{n}'_{obs},\unit{\varphi}')$ it follows that $\unit{\psi}=R(\xi,\unit{\varphi}')\unit{\psi}'\perp(\unit{n}_{obs},\unit{\varphi})$, and is in the direction of $\unit{\rho}_m$ on the observer's map.
We therefore get 
\begin{equation}
\begin{split}
\unit{P}_0= &-\left[\sin\theta'_{\rm b}\sin\left(\varphi'+\varphi'_{\rm b}\right)\right]\unit{\rho}\uum\\
&+\left[\cos\theta'\sin\theta'_{\rm b}\cos\left(\varphi'+\varphi'_{\rm b}\right)+\sin\theta'\cos\theta'_{\rm b}\right]\unit{\varphi}\uum.
\end{split}
\label{eq:pol_map}
\end{equation}
The polarization angle in each map cell is denoted by $\frac{P_{\varphi\uum}}{P_{\rho\uum}}$ and measured relative to %$\varphi_m=0$
 the local $\varphi\uum$:
\begin{equation}
\phi_{p_{_0}}=\varphi\uum+\tan^{-1}\left[\frac{\sin\theta'\cot\theta'_{\rm b}}{\sin\left(\varphi'+\varphi'_{\rm b}\right)}+\cos\theta'\cot\left(\varphi'+\varphi'_{\rm b}\right)\right].
\end{equation}
Transforming $\theta'$ to the lab frame results in
\begin{equation}
\begin{split}
\phi_{p_{_0}}&\simeq\varphi\uum\\&+\tan^{-1}\left[\frac{2y}{1+y^2}\frac{\cot\theta'_{\rm b}}{\sin\left(\varphi'+\varphi'_{\rm b}\right)}+\frac{1-y^2}{1+y^2}\cot\left(\varphi'+\varphi'_{\rm b}\right)\right],
\end{split}
\label{eq:theta_p}
\end{equation}
where $y=\Gamma\theta$.
This expression extends the analytic expressions obtained by \citet{Nava2015a} and by \citet{Granot2003a} for a uniform magnetic field on the plane of the shock ($\theta'_{\rm b}=\frac{\pi}{2}$, $\varphi'_{\rm b}=0$) and for a radial magnetic field ($\theta'_{\rm b}=0$) configurations. For an off-axis observer, $\xi$ is calculated in a similar way as we show in eq. \ref{eq:xi_rotation} and the rotation is done about the ($\unit{n}'\obs\times\unit{\beta}$) axis. The calculation of $\phi_{p_0}$ in this case is done numerically. 

%To calculate the total degree of polarization we decompose the polarization vector in each cell into two orthogonal components, the Stokes parameters
To calculate the total degree of polarization we construct the local Stokes parameters in each cell:
\begin{equation}
\begin{aligned}
	q_{_{\nu,0}}&=\Pi_\nu\cos2\phi_{p_{_0}},\\
	u_{_{\nu,0}}&=\Pi_\nu\sin2\phi_{p_{_0}}. \label{eq:LocalStokes}
\end{aligned}
\end{equation}
%Each parameter is weighted 
and weigh them by the surface brightness in the cell, divided by the total observed flux. The weighted $q_{_{\nu}}$ has the form
\begin{equation}
q_{_{\nu}}=\frac{q_{_{\nu,0}}\frac{dF_\nu}{dA}}{\int dF_{\nu}},
\end{equation}
%giving a total $Q$ parameter:
and is integrated to provide the total $Q$ parameter:
\begin{equation}
Q_{_{\nu}}=\int q_{_{\nu}} dA=\frac{\int q_{_{\nu,0}}\frac{dF_\nu}{dA}dA}{\int \frac{dF_\nu}{dA}dA},
\label{eq:Stocks_Q}
\end{equation}
where $\frac{dF_\nu}{dA}dA$ is given in eq. \ref{eq:dF_nu}. The parameter $U$ is calculated in the same manner.
From these we can obtain the global polarization degree, $P_{\nu}$ and the position angle of the total polarization vector, $\theta_{\nu,p}$:
 \begin{equation}
 \begin{aligned}
 P_{\nu}&=\sqrt{Q_{_\nu}^2+U_{_\nu}^2}\\
 \phi_{p,\nu}&=\frac{1}{2}\tan^{-1}\left(\frac{U_{_\nu}}{Q_{_\nu}}\right)\,. \label{eq:IntegratedPolDegAng}
 \end{aligned}
 \end{equation}
 Note that although the polarization angle in each cell is independent of frequency, the overall angle may depend on the observed frequency since the weight of each emitting region may depend on the frequency, as demonstrated in fig. \ref{fig:OnAxis_rand}.
%\begin{equation}
%    x=\frac{-b\pm\sqrt{b^2-4ac}}{2a}.
%	\label{eq:quadratic}
%\end{equation}
%
%Refer back to them as e.g. equation~(\ref{eq:quadratic}).

%%\subsection{Figures and tables}
%
%Figures and tables should be placed at logical positions in the text. Don't
%worry about the exact layout, which will be handled by the publishers.
%
%Figures are referred to as e.g. Fig.~\ref{fig:example_figure}, and tables as
%e.g. Table~\ref{tab:example_table}.
%
%% Example figure
%\begin{figure}
%	% To include a figure from a file named example.*
%	% Allowable file formats are eps or ps if compiling using latex
%	% or pdf, png, jpg if compiling using pdflatex
%	\includegraphics[width=\columnwidth]{example}
%    \caption{This is an example figure. Captions appear below each figure.
%	Give enough detail for the reader to understand what they're looking at,
%	but leave detailed discussion to the main body of the text.}
%    \label{fig:example_figure}
%\end{figure}
%
%% Example table
%\begin{table}
%	\centering
%	\caption{This is an example table. Captions appear above each table.
%	Remember to define the quantities, symbols and units used.}
%	\label{tab:example_table}
%	\begin{tabular}{lccr} % four columns, alignment for each
%		\hline
%		A & B & C & D\\
%		\hline
%		1 & 2 & 3 & 4\\
%		2 & 4 & 6 & 8\\
%		3 & 5 & 7 & 9\\
%		\hline
%	\end{tabular}
%\end{table}

\section{Results}
\label{sec:results}

We present results for polarization from a forward AG shock with a random magnetic field on the plane of the shock. We use the typical GRB parameters: $E_{\rm iso}=10^{52}$ ergs, $n=1$ cm$^{-3}$, $\theta_0=6^\circ$, $p=2.5$, $\varepsilon_e=0.1$, $\varepsilon_B=0.01$ and take a distance of $d_L=3.1\cdot10^{26}$ cm to the observer. 
In addition, the AG is modeled at times longer than $t_0$, thus we only use the slow cooling spectrum. We run simulations with the observer located at different viewing angles from the jet axis, parametrized by $q=\frac{\theta\obs}{\theta_0}$. 
%We run simulations where we vary the observer's viewing angle, parametrized by $q=\frac{\theta\obs}{\theta_0}$ and the observed frequency. 
We present the observed polarization in three characteristic frequencies: $10^{18}$ Hz for the X-ray band, typically above the cooling frequency; $10^{15}$ Hz for the optical band, typically between $\nu_c$ and $\nu_m$ and $10^{11}$ Hz for the microwave band, typically below the synchrotron frequency. 
%Since we do not model the self absorption frequency we do not run simularions in the radio band. 

\begin{figure*}	
	\includegraphics[width=\textwidth, trim=0cm 0cm 0cm 0cm]{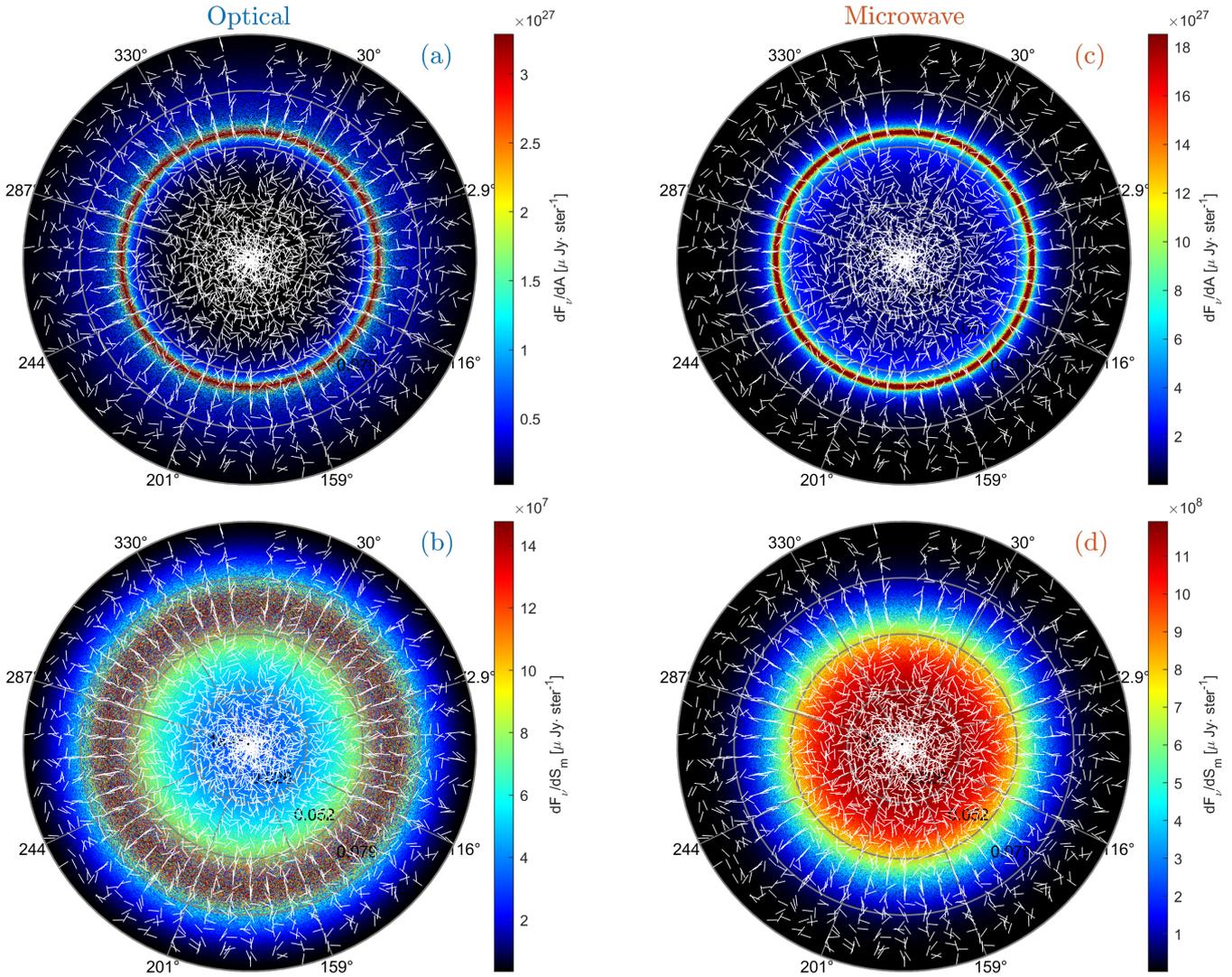}
	\caption{Intensity maps of an afterglow shock with the local polarization direction, projected on the observer's plane (short white lines). All maps are shown at the same observer time. 
	Left panels (a,b) show observed emission in the optical band ($\nu\obs>\nu_m$), and right panels (c,d) in the microwave band ($\nu\obs<\nu_m$). The top panels (a,c) show the surface brightness (flux per unit of solid angle on the plane of the sky). The surface brightness is dimmer at the center and diverges at the maximal opening angle of the EATS.
	The light at $\rho\uum<\theta\uuEATS$ originates from the front of the EATS while the light at $\rho\uum>\theta\uuEATS$ comes from the back of the EATS. The bottom panels show the flux per unit of surface on the map (corresponds to the flux per unit solid angle on the shock), which sets the scale of the polarized light at each point on the map. Here the observed image is different between the two bands. In the optical band most of the light is concentrated in a wide ring at opening angles close to $\theta\uuEATS$, while in the microwave the emission is strongest at the center. The reason for that is the different spectral slopes at each band as we explain in the text. A steeper slope in the optical band also results in a higher sensitivity to the angle of the local magnetic field from the LOS ($\sin\alpha'$) leading to a more granular image.
	The polarization is mostly radial at $\rho\uum$ close to $\theta\uuEATS$ and it has no preferred direction along the LOS. When the AG is observed off-axis, the optical image will show higher net polarization then the microwave, as we show in fig, \ref{fig:Together}. 
	}
	\label{fig:OnAxis_rand}
\end{figure*}

The observed polarization depends on the local polariation degree and on the spatial distribution of the observed light on the map. 
To demonstrate the effect of the light distribution on the total polarization,
%Qualitatively the observed polarization behaves differently at frequencies above and below $\nu_m$. To demonstrate that 
we show in fig. \ref{fig:OnAxis_rand} polarization maps in the optical (left) and microwave (right) bands seen by an observer aligned with the jet axis. The maps are 2D angular projection of a spherical-cap shock onto the observer plane, where the edge of the map corresponds to the jet opening angle, namely the maximal radial coordinate $\rho_{_{\rm m,max}}=\sin\theta_{0}\approx\theta_{0}$. The color scheme follows the observed intensity of light at each cell, and the short white lines mark the local direction of the polarization vector. %The white solid circle mark the opening angle of the emitting surface, $\theta\uuEATS$ (see fig. \ref{fig:shape_EATS}). 
All images are taken at the same $T\obs$.
The top panels (panels a,c) show the surface brightness, $\frac{dF_\nu}{dA}$, the flux per unit of observed solid angle on the plane of the sky. This quantity represents the intensity seen by an observer at each map cell. 
The bottom panels show the flux per unit solid angle on the map, $\frac{dF_\nu}{dS_m}$, which represents the weight to the polarized light in each map cell. Since each cell on the map matches a unique angle on the emitting surface, regions with large weight contribute most to the total polarization. 

\emph{Figure \ref{fig:OnAxis_rand} panels (a,c):} The surface brightness increases from the center, diverges at $\rho\uum=\theta\uuEATS$ and drops to zero at $\rho\uum>\theta\uuEATS$.
%(see Appendix B for a 1D plot of the light distribution). 
The region with $\rho\uum<\theta\uuEATS$ corresponds to the front of the EATS while $\rho\uum>\theta\uuEATS$ shows the back of the EATS.
(see fig. \ref{fig:shape_EATS} for clarification).
The surface brightness is the observed intensity scaled by a geometrical factor: the ratio of a differential solid angle on the emitting surface to its projection on the sky (eq. \ref{eq:dFdS_EATS}). 
The divergence of the surface brightness at $\theta\uuEATS$ occurs since the projected differential angle goes to zero and is a consequence of the 2D model. Another thing to notice is that the intensity of the microwave image is brighter than the optical at the center with respect to the peaks. 
The reason for that is the differences in the spectral slopes between the two bands. The observed intensity at cells with increasing $\rho\uum$ originates from regions on the EATS with higher $\Gamma$ and thus decreasing $\nu'$. Since $I'_{\nu'}$ has a positive (negative) slope in the microwave (optical) band, it becomes weaker (stronger) as $\rho\uum$ becomes larger. In the microwave band this effect counteracts the boost by the geometrical factor, leading to a more moderate increase in the surface brightness when moving from the center to the edge. As a result the center of the image appears brighter. The full calculation of the dependency of the surface brightness on $\rho\uum$ is shown in Appendix \ref{app:ComparisonToAnalytic}. 

%The top panels show the polarization on the shock surface in the lab frame (panels a,c), while the bottom panels show the polarization on the observer's map in the observer frame (panels c,d).
%The color schemes represent the observed intensity $I_\nu$ in a cell and the short white lines depict the local direction of the polarization vector. 
%All images are taken at the same $T\obs$.
\emph{Figure \ref{fig:OnAxis_rand} panel b:} Unlike the surface brightness, the differential flux density doesn't diverge at $\theta\uuEATS$. In the optical image it forms a wide ring of high intensity at angles close to $\theta\uuEATS$. The light coming from this ring is polarized mostly in the radial direction. To understand this, lets look at a circle with an opening angle $\theta\uuPOL=0.45/\Gamma\uuL=1/\Gamma(\theta\uuPOL)$ from the LOS. (see fig. \ref {fig:shape_EATS} for illustration and Appendix \ref{app:EATSEq} for the derivation). This angle is translated to $\theta\uuPOL'=\pi/2$ in the local emitting frame, namely the LOS is parallel to the shock surface and is aligned with the meridional direction ($\unit\theta'$). Suppose we take two orthogonal components of magnetic field in the local frame, an azimuthal component $b'_{\varphi'}$ and a meridional one $b'_{\theta'}$ of equal values. 
Since $b'_{\theta'}$ points in the direction of the LOS in the local frame, only radiation can be observed. When transforming back to the observer frame, the polarization vector rotates and points in the $\unit\rho\uum$ direction (see eq. \ref{eq:pol_map}). At the center of the map, the LOS is perpendicular to both $b'_{\varphi'}$ and $b'_{\theta'}$. Therefore the polarization vector doesn't have a preferred direction. Since the center is much dimmer than the polarized ring, the majority of observed light in the optical band is polarized in the radial direction.

\emph{Figure \ref{fig:OnAxis_rand} panel d:} 
The differential flux density in the microwave band behaves differently than in the optical band. Instead of being limb brightened, most of the light is concentrated at the center of the image where the polarization is low. This effect originates from the scaling of $I_\nu$ with $\rho\uum$, which was explained qualitatively above. The scaling of the flux density can be quantified from eq. \ref{eq:ObsIntensity} by noting that
\begin{equation}
dF_\nu\propto R^2I_\nu\propto D^{3-\kappa}R^3\Gamma^{1-3\kappa}\propto D^{3-\kappa}\Gamma^{-(1+3\kappa)},
\label{eq:dF_scaling}
\end{equation}
where $\kappa$ is the spectral slope at frequency $\nu$ and we ignore the contribution of $\sin\alpha'$ as it changes the result by a factor of order unity.
Substituting the values of $\Gamma$ and D at $\theta\uuEATS$ (see Appendix A for the exact values) and taking $\kappa={1/3}$,
it can be shown that the differential flux at $\theta\uuEATS$ is half the flux on the LOS. For this reason the total observed polarization at frequencies below $\nu_m$ is typically lower than the polarization at frequencies above $\nu_m$, as we show next.

\begin{figure*}
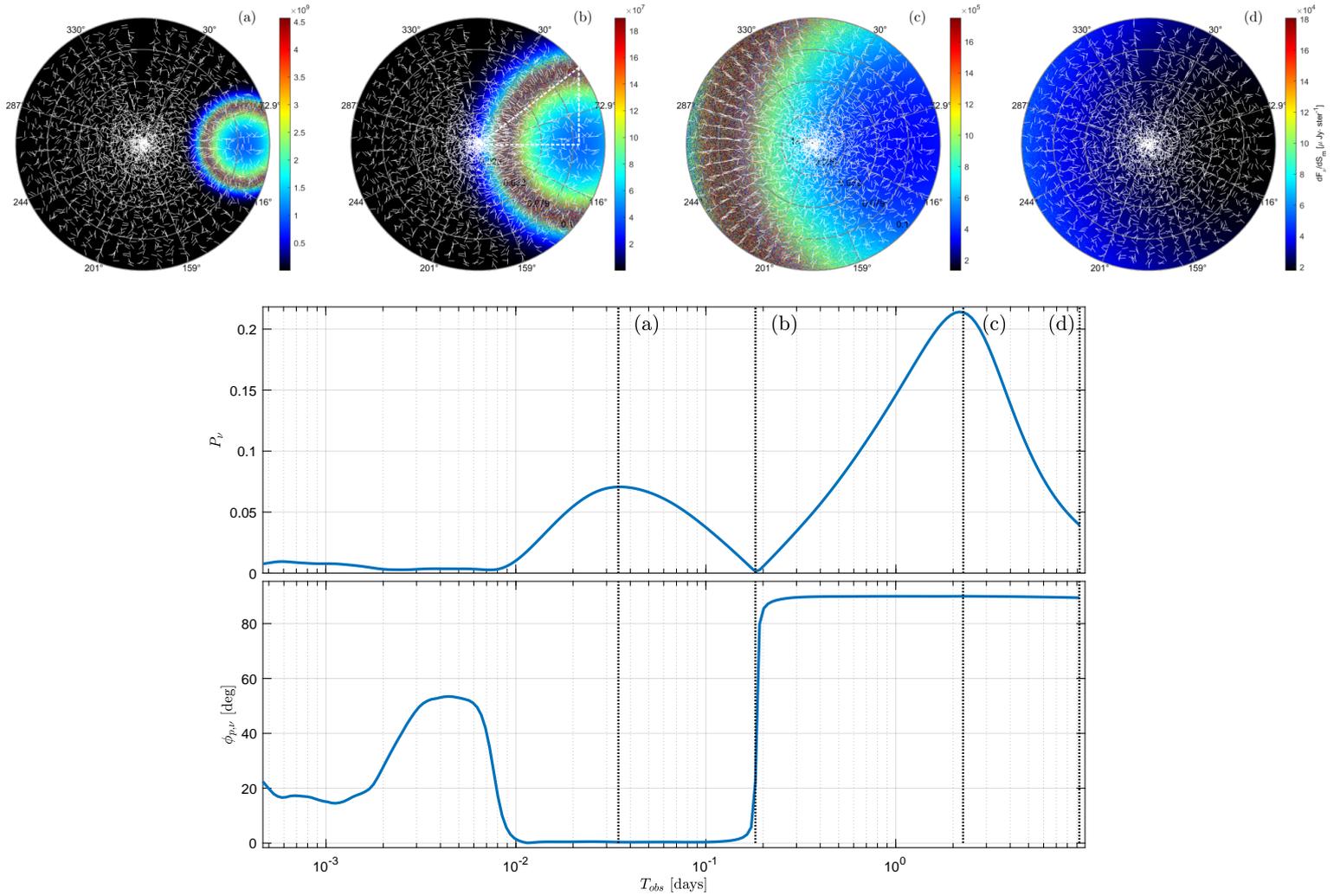
	
	\includegraphics[width=\textwidth, trim=3cm 0.5cm 3cm 0.5cm]{IMapsTogether7.png}
	\includegraphics[width=\textwidth, trim=0cm 3.0cm 0cm 2cm]{PAPDTogetherFinal.pdf}
	\caption{The time evolution of the total polarization in an AG shock observed at a frequency $\nu\obs=10^{15}$ Hz from an angle $\theta\obs=0.8\theta_0$. To focus on the geometrical effect alone, we set the local polarization degree $\Pi=0.75$ in all cells. Top panel: The differential flux per unit observer map, $\frac{dF}{dS\uum}$, with the local polarization direction shown in white lines. Middle panel: Time evolution of the integrated polarization degree. Bottom panel: The projected angle of the polarization vector on the map, measured from the vertical direction. In case b, when $P_\nu=0$ half of the emitting ring is visible. At this time the jet symmetry axis, the LOS and the point on the ring which touches the map edge form a right angle triangle, shown with dotted while line. the implications of that are discussed in section \ref{sec:Implications}}
	\label{fig:P2Together}
\end{figure*}

An AG shock with a random magnetic field shows no net polarization when viewed on-axis, since the observed ring is symmetric about the LOS and the polarization cancels out. 
When the AG is observed off-axis, the observed ring is no longer centered around the symmetry axis. As the shock slows down, the visible area of the shock grows, causing the ring to expand. Eventually some parts of the ring grow beyond the shock edge and disappear before others. This breaks the symmetry and leads to a net polarization with distinct features \citep[e.g.][]{Sari1999Pol,Ghisellini1999,Granot2003a}. 
Figure \ref{fig:P2Together} shows the evolution of an AG shock observed in the optical band at an angle $\theta\obs=0.8\theta_0$ from the symmetry axis ($q=0.8$).
%The evolution of the total polarization degree, $P_{\rm tot}$ is shown in the second panel. 
%We show here the polarization curves in the optical band.
%(blue) and microwave (red) bands. 
%We mark on each polarization curve four designated times with dashed lines, titled a-d. The observed polarization maps at each of these times are shown above (optical) and below (microwave) the $P_{\rm tot}$ panel, similar to fig. \ref{fig:OnAxis_rand}.At the bottom panel we show the evolution of the polarization vector position angle measured from the vertical direction. 
The middle panel shows the polarization curve and the bottom panel shows the evolution of the polarization vector position angle, measured from the vertical direction on the map. The dotted vertical lines mark four distinct episodes in the evolution of the polarization curve, titled a-d.
The polarization maps associated with each episode are shown at the top panel. 
The center of the emission ring in each map is at $\rho\uum=\theta\obs$ and is aligned with the LOS. 

The total polarization in fig. \ref{fig:P2Together} evolves as follows. As the forward shock decelerates, the 
ring expands and its right part disappears, creating a growing
deficit in the amount of light with horizontal polarization, thus the total polarization in the vertical direction increases. The polarization degree reaches its first peak when the inner radius of the ring touches the edge of the map  (\emph{case a}).
%Since the ring in the microwave band is smaller and emits light with a more isotropic polarization vector than the optical ring, the peak of the microwave bump is lower and occurs at a later time.  
As the ring continues to grow an increasing amount of light with predominantly vertical polarization is removed and the total polarization decreases. When $\sim1/2$ of the ring is outside the jet boundary, the vertically and horizontally polarized components balance each other and the net polarization zeros out (\emph{case b}).
From hereon, the dominant polarization component is horizontal and the total polarization position angle rotates by $90^{\circ}$. When the opening angle of the ring is $\sim(\theta_0+\theta\obs)$ the left side of the ring is at the edge of the jet, asymmetry reaches a maximum and the horizontal net polarization peaks (\emph{case c}). 
At longer times the ring disappears and the observed light originate from the inner, exceedingly dimmer parts of the ring, resulting in a steepening of the AG lightcurve associated with a jet-break. This implies that the maximal polarization and the jet-break should occur at times close to each other.
%This time is also occurs around the jet-break time when the AG lighcurve steepens. %Since the ring in the optical band is larger and more narrowly bounded around the maximally radially polarized light, the second peak occurs earlier and is higher than in the optical band.
%At longer times the ring disappears. The observed light comes from the inner ring parts where the polarization vector is more isotropically distributed and the net polarization drops (\emph{case d}).
Since the  polarization vector is more isotropically distributed in these parts, the net polarization drops (\emph{case d}).
In addition sideways expansion of the AG material, expected to occur after the jet-break (not simulated here), will lead to an even more symmetrical image and will reduce the total polarization degree even further \citep[see e.g.][]{Sari1999Pol}.
Altogether the evolution shows two distinct peaks between which the polarization vector rotates by $90^\circ$ and the polarization drops to zero. The occurrence time of these features and the height of the peaks depend on the observed frequency and on the viewing angle, as we show next. 

Note that a rotation of the polarization vector is seen in the optical band also at early times, when the net polarization is close to zero. At these times the entire ring is visible and the polarization cancels out almost completely. The non zero value of the position angle is obtained due to residuals in the polarization and is set arbitrarily. 
%Since it corresponds with zero polarization the it is insignificant. 

%When considering the X-ray band ($\nu>\nu_c$), the emission ring will be similar to the ring in the optical, however brighter with respect to the emission at the center of the image (eq. \ref{eq:dF_scaling}). Therefore the maximum polarization will be a bit higher and occur at the same time as in the optical band. In the microwave band ($\nu<\nu_m$) most of the emission comes from regions close to the LOS resulting in a lower polarization than in the optical.
%For the same viewing angle the rings inner radii will reach the map boundary at a later times than in the optical ring. 
%In addition, since the angular size of the microwave image is smaller then the optical or X-ray, the time when half of the image disappears will be delayed. Therefore the polarization zeroing time, which takes place during the rotation of the polarization vector will occur at a later time than in the optical and X-ray bands. Both these properties are seen in fig. \ref{fig:Together}.
%The second peak is related to the time when the image of polarized light attain maximum asymmetry. 
%is related to the jet-break time, when the  In the case of the radio light the radially polarized regions are dim at first however are becoming more illuminated at later times (ref for that?)
%therefore the peak will occur at later times.

\begin{figure*}
    \includegraphics[width=\textwidth, trim=2cm 3.5cm 2cm 4cm]{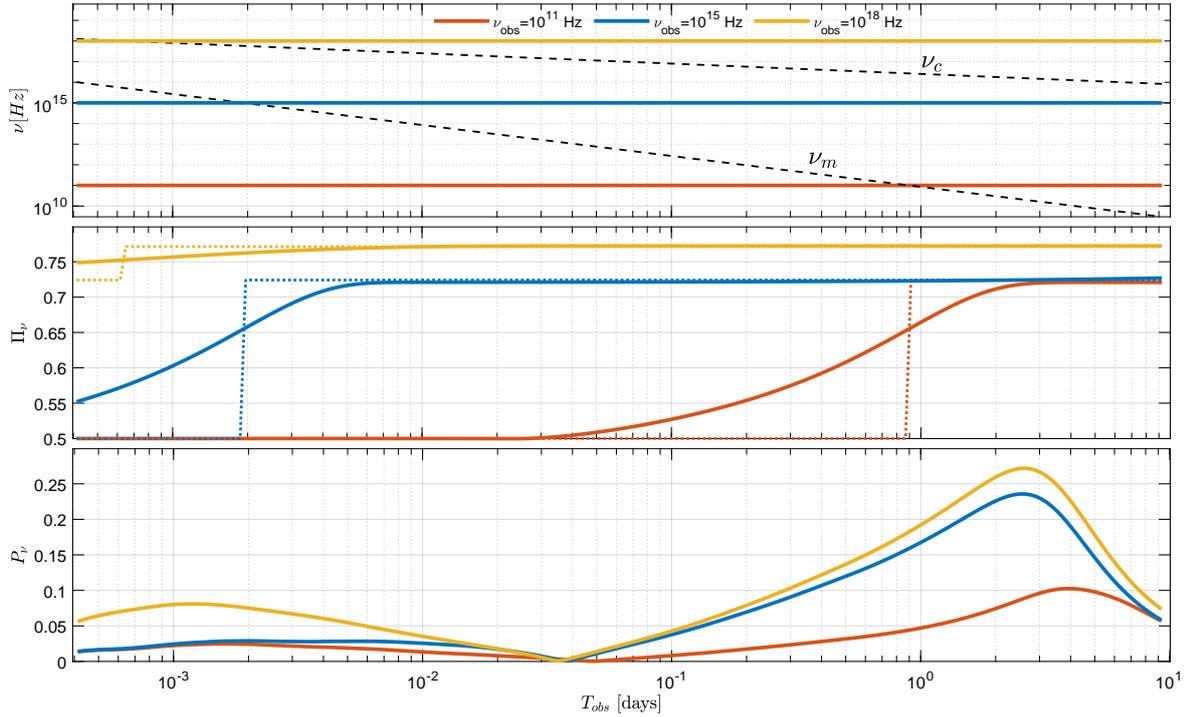}
    	\caption{Polarization curves at the three simulated bands: microwave (red), optical (blue) and X-ray (yellow) seen by an observer at $\theta\obs=0.95\theta_0$. 
    	Panel (a) shows the evolution of $\nu_c$ and $\nu_m$ on the LOS (dashed black lines) compared with the observing frequencies (solid colored lines).
    	Panel (b) shows the local polarization degree, $\Pi_\nu$ on the LOS. The solid and dotted lines show the value of $\Pi_\nu$ according to our numerical calculations and according to the analytic estimation of \citet{Granot2003} respectively.
    	Panel (c) shows the polarization curves. } 
	\label{fig:Together}
\end{figure*}

Apart from modifying the geometry of the observed image, the spectrum of the emitting particles also affects the polarization through the local polarization degree $\Pi_\nu$, as discussed in sec. \ref{ssec:PolSpect}. 
%Apart from geometrical effects, the overall polarization is affected by changes in the radiation spectrum as well. As discussed in sec. \ref{ssec:PolSpect}, the local polarization degree depends on the spectrum of the radiating particles. 
If the spectrum has a broken powerlaw shape, $\Pi_\nu$ will have different values depending on the the value of $\nu\obs$ relative to $\nu_m$ and $\nu_c$ (see fig.\ref{fig:PolSpectrum}). Since $\nu_m$ and $\nu_c$ decrease with time as the shock decelerates, they may cross $\nu\obs$. This crossing changes the spectral slope at $\nu\obs$  resulting in an 
increase in $\Pi_\nu$ and correspondingly in $P_{\nu}$.
Figure \ref{fig:Together} demonstrates this effect by showing the polarization curves seen in each of the three fiducial frequencies in microwave (red), optical (blue) and X-ray (yellow) bands. The observer is located at an angle of $\theta\obs=0.95\theta_0$ from the jet axis. 
Panel (a) shows the evolution of $\nu_m$ and $\nu_c$ in the observer frame along the LOS (dashed black lines). Panel (b) shows the evolution of $\Pi_\nu$ in each band along the LOS (solid lines). For comparison, we show in dotted lines the values of $\Pi_\nu$ based on the analytic spectral shape of a piecewise function \citet{Granot2003}.
Panel (c) shows the evolution of $P_{\nu}$, where the characteristic shape discussed above is seen in all three curves.
Looking at panel (a), at early times both the X-ray and optical frequencies are crossed by a critical frequency. The crossing has a larger effect on $\Pi_\nu$ in the optical band, increasing it by a factor of $1.5$
over a decade in the time scale (from a few times $10^{-4}$ days to $\sim5\times 10^{-3}$ days). This changes the shape of the first bump, stretching it to later times im comparison to the bumps of the other frequencies (panel c).
The peaks of the first bumps in the other wavelengths occur at roughly the same time. 
The ratio of the peak heights is directly related to the ratio of their $\Pi_\nu$ values, which is about 1.5. At a later time $\nu_m$ crosses the microwave frequency. The corresponding rise in $\Pi_\nu$ takes place throughout the entire duration of the second bump and pushes the peak of the bump to later times ($T\obs\sim4$ days) as opposed to 2.5 days in the optical and X-ray bands. The differences in the peak polarization here are related to the alignment of the polarization vectors in the emitting rings. In the optical and X-ray bands the emission rings are polarized in the radial direction, while in the microwave band the polarization vector is more isotropically oriented giving to a much lower overall polarization.

\begin{figure*}
	\includegraphics[width=\textwidth, trim=0cm 5.4cm 0cm 5.6cm]{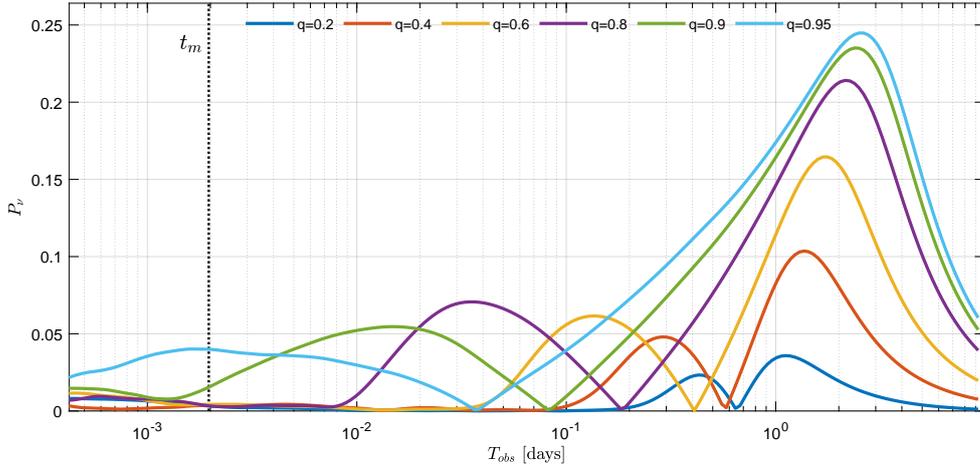}
	\caption{The time evolution of the integrated polarization degree at $\nu\obs=10^{15}$ Hz, measured by an observer at various viewing angles from the LOS (parameterized by $q=\frac{\theta\obs}{\theta_0}$). The dotted vertical line marks the time when $\nu_m$ drops below $\nu\obs$ resulting in an  increase in $P_{\nu}$ and to the stretch of the first bump in the polarization curve of $q=0.95$.}
	\label{fig:PDModels}
\end{figure*}

In figure \ref{fig:PDModels} we show the polarization curves in the optical band measured by observers with different $q$ values. %All curves show the same three geometrical features discussed above.
%in the context of fig. \ref{fig:OnAxis_rand}. 
The first peaks and zero polarization points occur at earlier times for larger $q$ values. This is consistent with the fact that both times are related to part of the emitting ring that remains inside the observed image and therefore are connected with $\theta\obs$, as shown in fig. \ref{fig:P2Together}. The stretching of the first bump
of the $q=0.95$ curve is a consequence of $\nu_m$ dropping below $\nu\obs$ at time $t_m$. 
The second peak occurs when the polarized ring is about to grow beyond the shock edge, which occurs when $(\theta_0+\theta\obs)\gtrsim\theta\uuPOL\simeq0.45/\Gamma\uuL$.
The condition gives a relation between $q$ and the time of the second peak, where $T\obs$ is roughly proportional to $(1+q)^{\frac{8}{3}}$, which explains the shift of the second peaks to later times at larger $q$ values. 
%For example, the ratio between the peak times of the polarization curves with $q=0.95$ and $q=0.5$ in fig. \ref{fig:PDModels} is $(2.6/1.4)\sim 2$ in agreement with the approximated expectation. 
%around the jet-break time in all curves with only a weak dependency on $q$. The height of the peak, however, does depend on $q$. %This is caused by increased asymmetry in the observed image. 
%Apart from the peak time, the peak polarization also depends on $q$.
The maximum polarization is connected with the asymmetry in the emission ring. As discussed above, the observed polarization at the time of the second peak is determined by the parts of the ring that remain in the observer's map (fig. \ref{fig:Together}c). 
At small $q$ values the ring center is close to the jet axis, and the ring image, which has a mean opening angle of $\sim\theta_0$ at that time, remains highly symmetric. As $q$ increases the ring center moves closer to the map edge, the asymmetry in the observed image increases and as a result the  polarization rises. The maximal values we obtain are consistent with other works \citep[e.g.][]{2004MNRAS.354...86R,Shimoda2020}.

\section{Implications on the viewing angle} \label{sec:Implications}
So far almost all measurements of linear polarization at times longer than a few thousand seconds show $P_{\nu}$ values smaller than $\sim5\%$ \citep[e.g.][]{2004ASPC..312..169C}. 
In those few GRBs where the time of the measured polarization coincides with a possible jet-break \citep[e.g.]{Greiner2003Natur,2012MNRAS.426....2W}, the polarization can be used to place an upper limit on the observer's viewing angle, since the jet-break occurs around the time of maximal polarization. Within our model the observed values imply  $q\lesssim0.2$.

If the magnetic field has a radial component as well, it will add an azimuthal polarization component that will reduce the total polarization and can account for higher values of $q$. In such a case there is an alternative way to constrain $\theta\obs$ by observing a zeroing of the polarization degree in coincidence with a rotation of the polarization angle by $90^\circ$. Such an event is indicative for a general configuration of random field with a uniform strength on the shock \citep[e.g.][]{Granot2003a}. 
The following is applicable for observations in the optical band. At the time of polarization zero, half of the emission ring is visible to the observer (see fig. \ref{fig:P2Together}b). At this time one can imagine a right-angle triangle between three points on the observer's map: the jet symmetry axis, the LOS and the point where the ring intersects with the map edge, 
%At this time the jet opening angle, the observer angle and the angle of maximal polarization in the ring, $\theta\uuPOL$, then relate to each other via  
which gives the relation 
\begin{equation}
\theta\obs^2+\theta_{_{\rm POL,1}}^2\simeq\theta_0^2. 
\end{equation}
If in addition a polarization peak is observed at a later time, or alternatively a jet-break is identified, one can get a second constraint on the system parameters,
%since at that time $\theta_0+\theta\obs\simeq\theta\uuEATS$ implying
\begin{equation}
\theta_0+\theta\obs\simeq\theta_{_{\rm POL,2}},
\label{eq:angle_calc2}
\end{equation}
where $\theta_{_{\rm POL,1}}$ and $\theta_{_{\rm POL,2}}$ are measured at the times of polarization zero and second polarization peak respectively. Note that we use $\theta\uuPOL$ rather than $\theta\uuEATS$ here, since the flux and the alignment of the polarization vector are maximal on $\theta\uuPOL$, thus it has the largest impact on the polarization evolution. 
Substituting the relation between $\theta\uuPOL$ and the Lorentz factor on the LOS, $\theta\uuPOL=0.45/\Gamma\uuL$ (see. appendix A), we get the expressions
\begin{equation}
 \begin{aligned}
 \theta_0&=\frac{0.45}{2\Gamma_{_{\rm L,2}}}\left(1+\left(\frac{\Gamma_{_{\rm L,2}}}{\Gamma_{_{\rm L,1}}}\right)^2\right)\\
 \theta\obs&=\frac{0.45}{2\Gamma_{_{\rm L,2}}}\left(1-\left(\frac{\Gamma_{_{\rm L,2}}}{\Gamma_{_{\rm L,1}}}\right)^2\right),
 \end{aligned}
 \label{eq:angles_calc}
\end{equation}
where $\Gamma_{_{\rm L,1}}$ and $\Gamma_{_{\rm L,2}}$ are the Lorentz factors associated with $\theta_{_{\rm POL,1}}$ and $\theta_{_{\rm POL,2}}$ respectively. Given an AG model, one can estimate $\Gamma\uuL(T\obs)$ (e.g. with eq. \ref{eq:Gamma_t}) and obtain both $\theta\obs$ and $\theta_0$. These relations work well with our simulation data, reproducing both angles at an accuracy of $\sim10\%$. %\gb{what do you mean by "these relations work well with our simulation data"? Aren't they calibrated by it so they should work well with it by default?}\obn{these are approximated relations and it is not clear how well they will work with the simulation data.}

The flip in the direction of the polarization vector, alongside with the zeroing of $P_\nu$, occurs almost instantaneously and requires continuous monitoring of the AG to be detected. To date we know of a single such event that was observed in GRB 021004 $\sim0.4$ days after the burst \citep{Rol2003A&A}. The AG was monitored by \citet{2003AJ....125.2291H} during the first month, who fitted it with $E_{_{\rm iso}}=(2.2\pm0.3)\times10^{52}$ erg, $n_{_{\rm ISM}}\approx35\times10^{-3}$ cm$^{-3}$ (assuming a uniform ambient density) and a jet-break time at $4.74\pm0.14$ days. Taking the jet break time as an indication for the polarization peak, calculating the associated Lorentz factors with eq. \ref{eq:Gamma_t} and substituting them in eq. \ref{eq:angles_calc} we obtain
%$\theta_0\simeq10^\circ$, $\theta\obs\simeq8^\circ$ giving $q\simeq0.8$. 
$\theta_0=10.3\pm0.3^\circ$,  $\theta\obs=7.5\pm0.3^\circ$ and $q=0.73\pm0.02$\footnote{The actual errors are likely larger due to  systematics, mostly the time of the peak polarization relative to the jet-break time.}. 
Our estimate of $\theta_0$ agrees with the one given by \citet{2003AJ....125.2291H}.
A relatively small polarization signature of $P_{\nu}\simeq1\%$ measured at $T\obs=3.5$ days, close to the jet-break \citep{Rol2003A&A}, indicates that 
%the magnetic field should contain both radial and tangential component of comparable strength.
there should be radial magnetic field in addition to the component tangential to the shock plane with a comparable strength.  %\gb{how can we say our estimates are correct if the magnetic field structure is different than ours? Maybe we should emphasise that a radial field will give a similar ring structure which will give similar relations?}\obn{a general random field will give the same pattern i discuss that in the previous paragraph}

%\subsection{Evolution of Completely Random MF}
%\subsection{Patchy Model}
%\subsection{GRB 121024A}
%\section{Discussion}
\section{Conclusions}
We present a computational method to calculate the observed polarization from relativistically moving surfaces by discretizing them, calculating the emission and polarization in each cell and summing up the flux weighted polarization from all cells to obtain the total polarization. 
Our method can work with arbitrary magnetic field configurations, system properties and observer viewing angles. It can also generate detailed polarization maps of the observed image that can help analyzing it.
We apply the method on AG forward shocks propagating in a medium with a uniform density and carrying random magnetic fields on the plane of the shock. We study the time evolution of the polarization vector observed at different frequencies relative to the synchrotron critical frequencies and different viewing angles.

We reproduce analytic results for the shape of the surface from which photons arrive to the observer simultaneously (EATS) and for the observed emission. %The surface brightness is limb brightened at the observed opening angle of the EATS, $\theta_{_{\rm EATS}}\simeq0.4/\Gamma_{_{\rm L}}$ \citep{Sari1997Egg}. 
The polarization at each cell in the observer's map is scaled by the flux through the cell. 
For $\nu\obs>\nu_m$ most of the flux is concentrated in a wide ring at the edge of the EATS having  
%For $\nu\obs>\nu_m$ the peak intensity in the ring has 
an angle $\theta\uuEATS\simeq0.4/\Gamma_{_{\rm L}}$ from the LOS \citep{Sari1997Egg}. The light coming from the ring is highly polarized in the radial direction, leading to a net polarization if the AG is observed off-axis and only part of the ring is revealed. 
In frequencies above $\nu_c$, the polarized ring is brighter due to the steeper spectral slope, leading to a higher polarization degree for the same conditions.
At frequencies below $\nu_m$ the situation is reversed. The flux is more evenly distributed on the map with the point along the LOS being twice as bright as at $\theta_{_{\rm EATS}}$, leading to a total polarization that is lower than in the previous cases. 
%In the optical band the radiation is narrowly concentrated around $\theta_{\rm peak}$ and is highly polarized in the radial direction. In the microwave and X-ray frequencies the ring is wider \citep{Granot2003a} and the polarization vector is more isotropically distributed. Thus higher polarization are expected in optical frequencies with respect to the other frequencies in this model.

We follow the time evolution of polarization in three frequencies where each frequency is associated with a different spectral regime in the the slow cooling synchrotron spectrum. All polarization curves viewed with $0<q<1$ show three distinct features of two bumps and a point of zero polarization in between where the polarization vector rotates by $90^\circ$ \citep[e.g.][]{Ghisellini1999,Sari1999Pol,Granot2003a,2004MNRAS.354...86R}. We show that in the optical band the first maximum occurs when the inner ring radius is touching the edge of the AG shock, and that the zero polarization occurs when half of the ring disappears. The second maximum occurs when the ring angular size is $\sim\theta_0+\theta\obs=\theta_0(1+q)\simeq\theta\uuPOL$, where the peak value increases with $q$. On $\theta\uuPOL$ the polarization vector is completely radial and has a maximal weight, thus this angle has the largest influence on the observed polarization.

The spectral slope also controls the value of the local polarization degree, $\Pi_\nu$. To properly account for this effect we calculated the specific intensity and the Stocks parameters (eqn. \ref{eq:ObsIntensity}, \ref{eq:Stocks_Q}), carefully modelling the spectral shape close to the critical frequencies, 
instead of using the standard analytic piece-wise approximation. We obtained smooth solutions for $\Pi_\nu$, which allowed us to better quantify the changes in the polarization degree when $\nu\obs$ is crossed by a critical frequency. We demonstrate the effect on the polarization curves in the three fiducial frequencies and show that the crossing leads to a gradual increase in the total polarization $P_{\nu}$, which can occur over more than an order of magnitude in time relative to the onset of the transition. This can 
alter the shape of bumps in polarization curves by increasing the value of the peak and pushing it to later times. 
%Crossing episodes can alter polarization curves in the optical band typically before the jet-break and in the microwave band at times close to the jet-break. 

Last, we introduced a method to estimate the jet opening angle and observing angle based on two distinct times, the time of zero polarization and the time of the second polarization peak, which occurs close to the jet-break time. We demonstrate the method on GRB 021004 and obtain a jet opening angle of $\theta_0\simeq10^\circ$, consistent with other estimations \citet[e.g.][]{2003AJ....125.2291H}, and an observed angle of $\theta\obs\simeq0.7\theta_0$. A polarization measurement of $P_\nu\sim1\%$ made at a time close to the jet-break time \citep{Rol2003A&A}, points to the existence of a magnetic field with
a radial component comparable to the component tangential to the shock. The conclusion agrees with the view that the magnetic field may develop a non negligible radial component downstream of the shock \citep[e.g.][]{2020MNRAS.491.5815G}. Since our model is 2D, the field we use is an average field from all emitting layers behind the shock. Additional detections of GRB AGs with large polarization angle rotations, accompanied by a zeroing of $P_\nu$, may help further constrain the observer viewing angle and the properties of magnetic field on the shock.

\section*{Acknowledgements}

We thank Ehud Nakar, Jonathan Granot and Ore Gottlieb for helpful discussions. This research was supported by an ISF grant 1657/18 and by an ISF (Icore) grant 1829/12. The maps in this work were plotted using the polarPcolor tool by \citet{https://doi.org/10.5281/zenodo.3774156}.

%%%%%%%%%%%%%%%%%%%%%%%%%%%%%%%%%%%%%%%%%%%%%%%%%%



\bibliographystyle{mnras}
\bibliography{ms.bbl} % if your bibtex file is called example.bib

\begin{thebibliography}{}
\makeatletter
\relax
\def\mn@urlcharsother{\let\do\@makeother \do\$\do\&\do\#\do\^\do\_\do\%\do\~}
\def\mn@doi{\begingroup\mn@urlcharsother \@ifnextchar [ {\mn@doi@}
  {\mn@doi@[]}}
\def\mn@doi@[#1]#2{\def\@tempa{#1}\ifx\@tempa\@empty \href
  {http://dx.doi.org/#2} {doi:#2}\else \href {http://dx.doi.org/#2} {#1}\fi
  \endgroup}
\def\mn@eprint#1#2{\mn@eprint@#1:#2::\@nil}
\def\mn@eprint@arXiv#1{\href {http://arxiv.org/abs/#1} {{\tt arXiv:#1}}}
\def\mn@eprint@dblp#1{\href {http://dblp.uni-trier.de/rec/bibtex/#1.xml}
  {dblp:#1}}
\def\mn@eprint@#1:#2:#3:#4\@nil{\def\@tempa {#1}\def\@tempb {#2}\def\@tempc
  {#3}\ifx \@tempc \@empty \let \@tempc \@tempb \let \@tempb \@tempa \fi \ifx
  \@tempb \@empty \def\@tempb {arXiv}\fi \@ifundefined
  {mn@eprint@\@tempb}{\@tempb:\@tempc}{\expandafter \expandafter \csname
  mn@eprint@\@tempb\endcsname \expandafter{\@tempc}}}

\bibitem[\protect\citeauthoryear{{Bersier} et~al.,}{{Bersier}
  et~al.}{2003}]{2003ApJ...583L..63B}
{Bersier} D.,  et~al., 2003, \mn@doi [\apjl] {10.1086/368158}, \href
  {https://ui.adsabs.harvard.edu/abs/2003ApJ...583L..63B} {583, L63}

\bibitem[\protect\citeauthoryear{{Biermann} \& {Cassinelli}}{{Biermann} \&
  {Cassinelli}}{1993}]{1993A&A...277..691B}
{Biermann} P.~L.,  {Cassinelli} J.~P.,  1993, \aap, \href
  {https://ui.adsabs.harvard.edu/abs/1993A&A...277..691B} {277, 691}

\bibitem[\protect\citeauthoryear{{Blandford} \& {McKee}}{{Blandford} \&
  {McKee}}{1976}]{1976PhFl...19.1130B}
{Blandford} R.~D.,  {McKee} C.~F.,  1976, \mn@doi [Physics of Fluids]
  {10.1063/1.861619}, \href
  {https://ui.adsabs.harvard.edu/abs/1976PhFl...19.1130B} {19, 1130}

\bibitem[\protect\citeauthoryear{Cheynet}{Cheynet}{2020}]{https://doi.org/10.5281/zenodo.3774156}
Cheynet E.,  2020, ECheynet/polarPcolor v3.8, \mn@doi{10.5281/ZENODO.3774156},
  \url {https://zenodo.org/record/3774156}

\bibitem[\protect\citeauthoryear{{Covino} et~al.,}{{Covino}
  et~al.}{1999a}]{1999GCN...330....1C}
{Covino} S.,  et~al., 1999a, GRB Coordinates Network, \href
  {https://ui.adsabs.harvard.edu/abs/1999GCN...330....1C} {330, 1}

\bibitem[\protect\citeauthoryear{{Covino} et~al.,}{{Covino}
  et~al.}{1999b}]{1999A&A...348L...1C}
{Covino} S.,  et~al., 1999b, \aap, \href
  {https://ui.adsabs.harvard.edu/abs/1999A&A...348L...1C} {348, L1}

\bibitem[\protect\citeauthoryear{{Covino} et~al.,}{{Covino}
  et~al.}{2003}]{2003A&A...400L...9C}
{Covino} S.,  et~al., 2003, \mn@doi [\aap] {10.1051/0004-6361:20030133}, \href
  {https://ui.adsabs.harvard.edu/abs/2003A&A...400L...9C} {400, L9}

\bibitem[\protect\citeauthoryear{{Covino}, {Ghisellini}, {Lazzati}  \&
  {Malesani}}{{Covino} et~al.}{2004}]{2004ASPC..312..169C}
{Covino} S.,  {Ghisellini} G.,  {Lazzati} D.,   {Malesani} D.,  2004, in
  {Feroci} M.,  {Frontera} F.,  {Masetti} N.,   {Piro} L.,  eds,  Astronomical
  Society of the Pacific Conference Series Vol. 312, Gamma-Ray Bursts in the
  Afterglow Era. p.~169 (\mn@eprint {arXiv} {astro-ph/0301608})

\bibitem[\protect\citeauthoryear{{Ghisellini} \& {Lazzati}}{{Ghisellini} \&
  {Lazzati}}{1999}]{Ghisellini1999}
{Ghisellini} G.,  {Lazzati} D.,  1999, \mn@doi [\mnras]
  {10.1046/j.1365-8711.1999.03025.x}, \href
  {https://ui.adsabs.harvard.edu/abs/1999MNRAS.309L...7G} {309, L7}

\bibitem[\protect\citeauthoryear{{Gill} \& {Granot}}{{Gill} \&
  {Granot}}{2020}]{2020MNRAS.491.5815G}
{Gill} R.,  {Granot} J.,  2020, \mn@doi [\mnras] {10.1093/mnras/stz3340}, \href
  {https://ui.adsabs.harvard.edu/abs/2020MNRAS.491.5815G} {491, 5815}

\bibitem[\protect\citeauthoryear{{Gill}, {Granot}  \& {Kumar}}{{Gill}
  et~al.}{2019}]{2019MNRAS.tmp.2582G}
{Gill} R.,  {Granot} J.,   {Kumar} P.,  2019, \mn@doi [\mnras]
  {10.1093/mnras/stz2976}, \href
  {https://ui.adsabs.harvard.edu/abs/2019MNRAS.tmp.2582G} {p.~2582}

\bibitem[\protect\citeauthoryear{Granot}{Granot}{2003}]{Granot2003}
Granot J.,  2003, \mn@doi [The Astrophysical Journal] {10.1086/379110}, 596,
  L17

\bibitem[\protect\citeauthoryear{{Granot}}{{Granot}}{2008}]{2008MNRAS.390L..46G}
{Granot} J.,  2008, \mn@doi [\mnras] {10.1111/j.1745-3933.2008.00533.x}, \href
  {https://ui.adsabs.harvard.edu/abs/2008MNRAS.390L..46G} {390, L46}

\bibitem[\protect\citeauthoryear{Granot \& Konigl}{Granot \&
  Konigl}{2003}]{Granot2003a}
Granot J.,  Konigl A.,  2003, \mn@doi [The Astrophysical Journal]
  {10.1086/378733}, 594, L83

\bibitem[\protect\citeauthoryear{{Granot}, {Piran}  \& {Sari}}{{Granot}
  et~al.}{1999}]{Granot1999Apj}
{Granot} J.,  {Piran} T.,   {Sari} R.,  1999, \mn@doi [\apj] {10.1086/306884},
  \href {https://ui.adsabs.harvard.edu/abs/1999ApJ...513..679G} {513, 679}

\bibitem[\protect\citeauthoryear{{Greiner} et~al.,}{{Greiner}
  et~al.}{2003}]{Greiner2003Natur}
{Greiner} J.,  et~al., 2003, \mn@doi [\nat] {10.1038/nature02077}, \href
  {https://ui.adsabs.harvard.edu/abs/2003Natur.426..157G} {426, 157}

\bibitem[\protect\citeauthoryear{{Gruzinov} \& {Waxman}}{{Gruzinov} \&
  {Waxman}}{1999}]{1999ApJ...511..852G}
{Gruzinov} A.,  {Waxman} E.,  1999, \mn@doi [\apj] {10.1086/306720}, \href
  {https://ui.adsabs.harvard.edu/abs/1999ApJ...511..852G} {511, 852}

\bibitem[\protect\citeauthoryear{{Holland} et~al.,}{{Holland}
  et~al.}{2003}]{2003AJ....125.2291H}
{Holland} S.~T.,  et~al., 2003, \mn@doi [\aj] {10.1086/374235}, \href
  {https://ui.adsabs.harvard.edu/abs/2003AJ....125.2291H} {125, 2291}

\bibitem[\protect\citeauthoryear{{Katz}}{{Katz}}{1994}]{1994ApJ...422..248K}
{Katz} J.~I.,  1994, \mn@doi [\apj] {10.1086/173723}, \href
  {https://ui.adsabs.harvard.edu/abs/1994ApJ...422..248K} {422, 248}

\bibitem[\protect\citeauthoryear{{Katz} \& {Piran}}{{Katz} \&
  {Piran}}{1997}]{1997ApJ...490..772K}
{Katz} J.~I.,  {Piran} T.,  1997, \mn@doi [\apj] {10.1086/304913}, \href
  {https://ui.adsabs.harvard.edu/abs/1997ApJ...490..772K} {490, 772}

\bibitem[\protect\citeauthoryear{{Klose}, {Palazzi}, {Masetti}, {Stecklum},
  {Greiner}, {Hartmann}  \& {Schmid}}{{Klose} et~al.}{2004}]{Klose2004A&A}
{Klose} S.,  {Palazzi} E.,  {Masetti} N.,  {Stecklum} B.,  {Greiner} J.,
  {Hartmann} D.~H.,   {Schmid} H.~M.,  2004, \mn@doi [\aap]
  {10.1051/0004-6361:20041024}, \href
  {https://ui.adsabs.harvard.edu/abs/2004A&A...420..899K} {420, 899}

\bibitem[\protect\citeauthoryear{{Kobayashi}, {Piran}  \& {Sari}}{{Kobayashi}
  et~al.}{1999}]{1999ApJ...513..669K}
{Kobayashi} S.,  {Piran} T.,   {Sari} R.,  1999, \mn@doi [\apj]
  {10.1086/306868}, \href
  {https://ui.adsabs.harvard.edu/abs/1999ApJ...513..669K} {513, 669}

\bibitem[\protect\citeauthoryear{{Laskar} et~al.,}{{Laskar}
  et~al.}{2019}]{Laskar2019ApJ}
{Laskar} T.,  et~al., 2019, \mn@doi [\apjl] {10.3847/2041-8213/ab2247}, \href
  {https://ui.adsabs.harvard.edu/abs/2019ApJ...878L..26L} {878, L26}

\bibitem[\protect\citeauthoryear{{Medvedev} \& {Loeb}}{{Medvedev} \&
  {Loeb}}{1999}]{1999ApJ...526..697M}
{Medvedev} M.~V.,  {Loeb} A.,  1999, \mn@doi [\apj] {10.1086/308038}, \href
  {https://ui.adsabs.harvard.edu/abs/1999ApJ...526..697M} {526, 697}

\bibitem[\protect\citeauthoryear{{Medvedev}, {Fiore}, {Fonseca}, {Silva}  \&
  {Mori}}{{Medvedev} et~al.}{2005}]{2005ApJ...618L..75M}
{Medvedev} M.~V.,  {Fiore} M.,  {Fonseca} R.~A.,  {Silva} L.~O.,   {Mori}
  W.~B.,  2005, \mn@doi [\apjl] {10.1086/427921}, \href
  {https://ui.adsabs.harvard.edu/abs/2005ApJ...618L..75M} {618, L75}

\bibitem[\protect\citeauthoryear{{M{\'e}sz{\'a}ros}, {Rees}  \&
  {Wijers}}{{M{\'e}sz{\'a}ros} et~al.}{1998}]{1998ApJ...499..301M}
{M{\'e}sz{\'a}ros} P.,  {Rees} M.~J.,   {Wijers} R.~A.~M.~J.,  1998, \mn@doi
  [\apj] {10.1086/305635}, \href
  {https://ui.adsabs.harvard.edu/abs/1998ApJ...499..301M} {499, 301}

\bibitem[\protect\citeauthoryear{{Nakar} \& {Oren}}{{Nakar} \&
  {Oren}}{2004}]{2004ApJ...602L..97N}
{Nakar} E.,  {Oren} Y.,  2004, \mn@doi [\apjl] {10.1086/382729}, \href
  {https://ui.adsabs.harvard.edu/abs/2004ApJ...602L..97N} {602, L97}

\bibitem[\protect\citeauthoryear{Nava, Nakar  \& Piran}{Nava
  et~al.}{2015}]{Nava2015a}
Nava L.,  Nakar E.,   Piran T.,  2015, \mn@doi [MNRAS]
  {https://doi.org/10.1093/mnras/stv2434nras/stv2434}, 1606, 1594

\bibitem[\protect\citeauthoryear{{Paczynski} \& {Rhoads}}{{Paczynski} \&
  {Rhoads}}{1993}]{1993ApJ...418L...5P}
{Paczynski} B.,  {Rhoads} J.~E.,  1993, \mn@doi [\apjl] {10.1086/187102}, \href
  {https://ui.adsabs.harvard.edu/abs/1993ApJ...418L...5P} {418, L5}

\bibitem[\protect\citeauthoryear{{Piran}, {Shemi}  \& {Narayan}}{{Piran}
  et~al.}{1993}]{1993MNRAS.263..861P}
{Piran} T.,  {Shemi} A.,   {Narayan} R.,  1993, \mn@doi [\mnras]
  {10.1093/mnras/263.4.861}, \href
  {https://ui.adsabs.harvard.edu/abs/1993MNRAS.263..861P} {263, 861}

\bibitem[\protect\citeauthoryear{{Planck Collaboration} et~al.,}{{Planck
  Collaboration} et~al.}{2018}]{2018A&A...610C...1P}
{Planck Collaboration} et~al., 2018, \mn@doi [\aap]
  {10.1051/0004-6361/201322612e}, \href
  {https://ui.adsabs.harvard.edu/abs/2018A&A...610C...1P} {610, C1}

\bibitem[\protect\citeauthoryear{{Rol} et~al.,}{{Rol}
  et~al.}{2000}]{2000ApJ...544..707R}
{Rol} E.,  et~al., 2000, \mn@doi [\apj] {10.1086/317256}, \href
  {https://ui.adsabs.harvard.edu/abs/2000ApJ...544..707R} {544, 707}

\bibitem[\protect\citeauthoryear{{Rol} et~al.,}{{Rol}
  et~al.}{2003}]{Rol2003A&A}
{Rol} E.,  et~al., 2003, \mn@doi [\aap] {10.1051/0004-6361:20030731}, \href
  {https://ui.adsabs.harvard.edu/abs/2003A&A...405L..23R} {405, L23}

\bibitem[\protect\citeauthoryear{{Rossi}, {Lazzati}, {Salmonson}  \&
  {Ghisellini}}{{Rossi} et~al.}{2004}]{2004MNRAS.354...86R}
{Rossi} E.~M.,  {Lazzati} D.,  {Salmonson} J.~D.,   {Ghisellini} G.,  2004,
  \mn@doi [\mnras] {10.1111/j.1365-2966.2004.08165.x}, \href
  {https://ui.adsabs.harvard.edu/abs/2004MNRAS.354...86R} {354, 86}

\bibitem[\protect\citeauthoryear{Rybicki \& Lightman}{Rybicki \&
  Lightman}{1979}]{Rybicki1979}
Rybicki G.~B.,  Lightman A.~P.,  1979, {Radiative processes in astrophysics}.
New York : Wiley, New York

\bibitem[\protect\citeauthoryear{{Sari}}{{Sari}}{1997}]{Sari1997Hydro}
{Sari} R.,  1997, \mn@doi [\apjl] {10.1086/310957}, \href
  {https://ui.adsabs.harvard.edu/abs/1997ApJ...489L..37S} {489, L37}

\bibitem[\protect\citeauthoryear{{Sari}}{{Sari}}{1998}]{Sari1997Egg}
{Sari} R.,  1998, \mn@doi [\apjl] {10.1086/311160}, \href
  {https://ui.adsabs.harvard.edu/abs/1998ApJ...494L..49S} {494, L49}

\bibitem[\protect\citeauthoryear{Sari}{Sari}{1999a}]{Sari1999Pol}
Sari R.,  1999a, \mn@doi [The Astrophysical Journal] {10.1086/312294}, 524, L43

\bibitem[\protect\citeauthoryear{{Sari}}{{Sari}}{1999b}]{1999ApJ...524L..43S}
{Sari} R.,  1999b, \mn@doi [\apjl] {10.1086/312294}, \href
  {https://ui.adsabs.harvard.edu/abs/1999ApJ...524L..43S} {524, L43}

\bibitem[\protect\citeauthoryear{{Sari}, {Piran}  \& {Narayan}}{{Sari}
  et~al.}{1998}]{Sari1998}
{Sari} R.,  {Piran} T.,   {Narayan} R.,  1998, \mn@doi [\apjl]
  {10.1086/311269}, \href
  {https://ui.adsabs.harvard.edu/abs/1998ApJ...497L..17S} {497, L17}

\bibitem[\protect\citeauthoryear{{Shimoda} \& {Toma}}{{Shimoda} \&
  {Toma}}{2020}]{Shimoda2020}
{Shimoda} J.,  {Toma} K.,  2020, arXiv e-prints, \href
  {https://ui.adsabs.harvard.edu/abs/2020arXiv200503710S} {p. arXiv:2005.03710}

\bibitem[\protect\citeauthoryear{{Waxman}}{{Waxman}}{1997a}]{1997ApJ...485L...5W}
{Waxman} E.,  1997a, \mn@doi [\apjl] {10.1086/310809}, \href
  {https://ui.adsabs.harvard.edu/abs/1997ApJ...485L...5W} {485, L5}

\bibitem[\protect\citeauthoryear{{Waxman}}{{Waxman}}{1997b}]{1997ApJ...489L..33W}
{Waxman} E.,  1997b, \mn@doi [\apjl] {10.1086/310960}, \href
  {https://ui.adsabs.harvard.edu/abs/1997ApJ...489L..33W} {489, L33}

\bibitem[\protect\citeauthoryear{{Wiersema} et~al.,}{{Wiersema}
  et~al.}{2012}]{2012MNRAS.426....2W}
{Wiersema} K.,  et~al., 2012, \mn@doi [\mnras]
  {10.1111/j.1365-2966.2012.20943.x}, \href
  {https://ui.adsabs.harvard.edu/abs/2012MNRAS.426....2W} {426, 2}

\bibitem[\protect\citeauthoryear{{Wijers} et~al.,}{{Wijers}
  et~al.}{1999}]{1999ApJ...523L..33W}
{Wijers} R.~A.~M.~J.,  et~al., 1999, \mn@doi [\apjl] {10.1086/312262}, \href
  {https://ui.adsabs.harvard.edu/abs/1999ApJ...523L..33W} {523, L33}

\makeatother
\end{thebibliography}


\begin{thebibliography}{}
\makeatletter
\relax
\def\mn@urlcharsother{\let\do\@makeother \do\$\do\&\do\#\do\^\do\_\do\%\do\~}
\def\mn@doi{\begingroup\mn@urlcharsother \@ifnextchar [ {\mn@doi@}
  {\mn@doi@[]}}
\def\mn@doi@[#1]#2{\def\@tempa{#1}\ifx\@tempa\@empty \href
  {http://dx.doi.org/#2} {doi:#2}\else \href {http://dx.doi.org/#2} {#1}\fi
  \endgroup}
\def\mn@eprint#1#2{\mn@eprint@#1:#2::\@nil}
\def\mn@eprint@arXiv#1{\href {http://arxiv.org/abs/#1} {{\tt arXiv:#1}}}
\def\mn@eprint@dblp#1{\href {http://dblp.uni-trier.de/rec/bibtex/#1.xml}
  {dblp:#1}}
\def\mn@eprint@#1:#2:#3:#4\@nil{\def\@tempa {#1}\def\@tempb {#2}\def\@tempc
  {#3}\ifx \@tempc \@empty \let \@tempc \@tempb \let \@tempb \@tempa \fi \ifx
  \@tempb \@empty \def\@tempb {arXiv}\fi \@ifundefined
  {mn@eprint@\@tempb}{\@tempb:\@tempc}{\expandafter \expandafter \csname
  mn@eprint@\@tempb\endcsname \expandafter{\@tempc}}}

\makeatother
\end{thebibliography}

%\begin{thebibliography}{99}
%\end{thebibliography}

%%%%%%%%%%%%%%%%%%%%%%%%%%%%%%%%%%%%%%%%%%%%%%%%%%

%%%%%%%%%%%%%%%%% APPENDICES %%%%%%%%%%%%%%%%%%%%%
\
\appendix
\section{The EATS equations}
\label{app:EATSEq} 
In order to obtain the EATS quantities used in this work we redevelop the EATS equations from \citep{Sari1997Egg} and derive various quantities used in this work. We start with the equation for the radius of the EATS, eq. \ref{eq:EAT_surface}:
\begin{equation}
R(T\obs,\mu\obs)=\frac{cT\obs}{1-\mu\obs + \frac{1}{16\Gamma^2}},
\label{App:R_EATS}
\end{equation}
where $\Gamma$ is the Lorentz factor of the fluid just behind the shock. We can express it in terms of the Lorentz factor and EATS radius on the LOS through $\Gamma=\Gamma\uuL\left(\frac{R}{R\uuL}\right)^{-\frac{3}{2}}$, where
$R\uuL=16\Gamma\uuL^2T\obs$. Substituting these in eq. \ref{App:R_EATS} we get an equation for the observed EATS opening angle  
\begin{equation}
1-\mu\obs=\frac{1}{16\Gamma\uuL^2}\left(\frac{R\uuL}{R}-\left(\frac{R}{R\uuL}\right)^3\right),
\label{App:mu_EATS}
\end{equation}
and from that we can obtain the perpendicular radius to the LOS at each point:
\begin{equation}
R_\perp=R\sqrt{1-\mu\obs^2}=\frac{\sqrt{2}R\uuL}{4\Gamma\uuL}\sqrt{\frac{R}{R\uuL}-\left(\frac{R}{R\uuL}\right)^5}.
\label{App:Rperp_EATS}
\end{equation}
The surface brightness is defined as the specific flux per unit of solid angle on the sky, 
\begin{equation}
\frac{dF_\nu(\rho\uum,\varphi\uum)}{dA}=I_\nu\left(\frac{R}{d_{_{\rm L}}}\right)^2\mu\obs\frac{d_{_{\rm L}}^2}{R_\perp}\frac{d\mu\obs}{dR_\perp},
\label{App:dFdA}
\end{equation}
where $dA=\frac{R_\perp dR_\perp d\varphi}{d\uuL^2}$ and $d\uuL$ is the distance from the source. To calculate this quantity we need to evaluate 
$\frac{d\mu\obs}{dR\perp}=\frac{d\mu\obs}{dR}\frac{dR}{dR_\perp}$.
From eq. \ref{App:mu_EATS} we get that
\begin{equation}
\frac{d\mu\obs}{dR}=\frac{R\uuL}{16\Gamma\uuL^2R^2}\left(1+3\left(\frac{R}{R\uuL}\right)^4\right),
\end{equation}
and from eq. \ref{App:Rperp_EATS}:
\begin{equation}
\frac{dR_\perp}{dR}=\frac{R\uuL}{16\Gamma\uuL^2R_\perp}\left(1-5\left(\frac{R}{R\uuL}\right)^4\right)
\end{equation}
resulting in:
\begin{equation}
\frac{1}{R_\perp}\frac{d\mu\obs}{dR_\perp}=\frac{1}{R^2}\frac{\left(1+3\left(\frac{R}{R\uuL}\right)^4\right)}{\left(1-5\left(\frac{R}{R\uuL}\right)^4\right)}.
\label{App:Ang_diff_EATS}
\end{equation}
Eq. \ref{App:Ang_diff_EATS} can be substituted in eq. \ref{App:dFdA} to obtain the surface brightness 
\begin{equation}
\frac{dF_\nu(\rho\uum,\varphi\uum)}{dA}=I_\nu\mu\obs\frac{\left(1+3\left(\frac{R}{R\uuL}\right)^4\right)}{\left(1-5\left(\frac{R}{R\uuL}\right)^4\right)},
\label{App:dFdA2}
\end{equation}
which is given in eq. \ref{eq:dFdS}. We further evaluate the opening angle of the EATS by equating $\frac{dR_\perp}{dR}=0$ giving 
\begin{equation}
R_{_{\rm EATS}}=R\uuL\left(\frac{1}{5}\right)^{\frac{1}{4}},
\end{equation}
and substituting that in eq. \ref{App:Rperp_EATS}:
\begin{equation}
R_{_{\perp,\rm EATS}}=\frac{R\uuL}{\Gamma\uuL}\frac{1}{5^{\frac{1}{8}}\sqrt{10}}
\end{equation}
The EATS opening angle is defined as $\theta\uuEATS\simeq\sin\theta_{_{\rm EATS}}=\frac{R_{_{\perp,\rm EATS}}}{R_{_{\rm EATS}}}$:
\begin{equation}
\theta\uuEATS\simeq\frac{5^{\frac{1}{8}}}{\sqrt{10}}\frac{1}{\Gamma\uuL}=\frac{1}{\sqrt{2}\Gamma_{_{\rm EATS}}}.
\end{equation}
Both $R_{_{\perp,\rm EATS}}$ and $\theta\uuEATS$ are shown in fig. \ref{eq:EAT_surface}. Last, we derive the angle $\theta\uuPOL$ on the EATS, for which $\Gamma\uuPOL\theta\uuPOL=1$. At this angle the observed polarization is purely radial. For this we solve the equation 
\begin{equation}
\frac{R_\perp}{R} \Gamma\uuL\left(\frac{R\uuL}{R}\right)^{\frac{3}{2}} =\frac{\sqrt{2}}{4}\left(\frac{R\uuL}{R}\right)^2\sqrt{1-\left(\frac{R}{R\uuL}\right)^4}=1,
\end{equation}
resulting in
\begin{equation}
\begin{split}
&R\uuPOL=R\uuL\left(\frac{1}{9}\right)^{\frac{1}{4}}\\
&R_{_{\perp,\rm POL}}=\frac{R\uuL}{\Gamma\uuL}\frac{1}{9^{\frac{1}{8}}\sqrt{9}},
\end{split}
\end{equation}
and a corresponding opening angle
\begin{equation}
\theta\uuPOL\simeq\frac{9^{\frac{1}{8}}}{\sqrt{9}}\frac{1}{\Gamma\uuL}=\frac{1}{\Gamma_{_{\rm POL}}}.
\end{equation}
Note that $\theta\uuPOL$ is slightly larger than $\theta\uuEATS$ and is located at the back part of the EATS.

\section{Comparisons to analytic results}
\label{app:ComparisonToAnalytic} 

\begin{figure}
    \includegraphics[width=\columnwidth, trim=0cm 0cm 0cm 0cm]{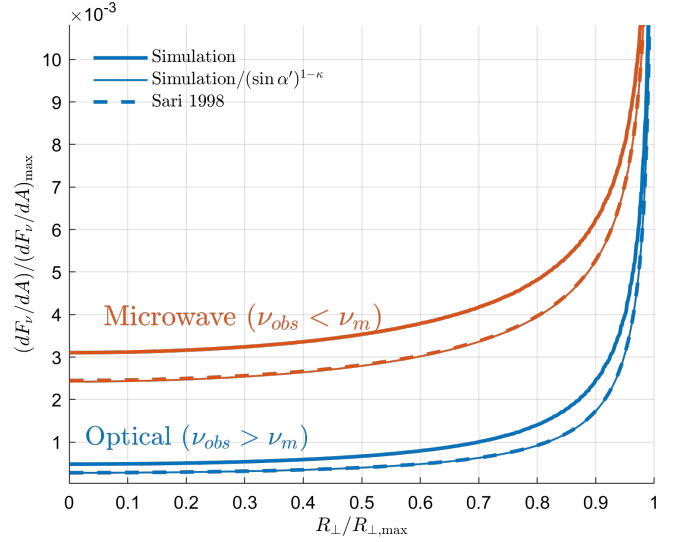}
    	\caption{Normalized surface brightness as function of $R_{\perp}$ for two different observed frequencies - optical (blue lines, $\nu_{\obs}>\nu_m$) and microwave (red lines, $\nu_{\obs}<\nu_m$). The results of our simulations at $q=0$ are shown in solid lines while the analytical expressions of \citet{Sari1997Egg} are in dashed lines.
    	}
	\label{fig:SurfaceBrightnessTest}
\end{figure}

We test the angular dependency of the surface brightness on the EATS, by comparing it to the analytic expressions in \citet{Sari1997Egg}.
An important difference between our method and the analytic calculation is how each method considers the effect of the pitch angle, $\alpha'$, on the observed emission.
The synchrotron power of an electron at an angle $\theta$ from the LOS
depends on $(\sin\alpha')$, where $\alpha'=\cos^{-1}({\unit{b}'}\cdot\unit{\theta}')$.
Analytic methods have a hard time calculating  $\sin\alpha'$ in a random field, since $\unit{b}'$ in each point is not defined. Instead they use the averaged value on a sphere $\langle\sin\alpha'\rangle=\pi/4$, which is independent on the location on the EATS. We assume a uniform field at each cell with a random direction, thus the code can calculate the actual value of $\sin\alpha'$ at each point. To compare the angular dependency of the simulation output with the 1D analytic expression we divide the intensity at each cell by $(\sin\alpha')^{1-\kappa}$,  
where $\kappa$ is the local spectral index (see eq. \ref{eq:ObsIntensity}) and average the result over the $\unit{\phi'}$ direction. 
Figure \ref{fig:SurfaceBrightnessTest} shows 1D curves of the surface brigtness as a function of $R_{\perp}$ (eq. \ref{App:Rperp_EATS}).
We show curves for the optical (blue) and microwave (red) frequencies. 
Note that we only show the curves from the front of the EATS ($\theta<\theta\uuEATS$).
The dashed lines show the analytic curves. The thin solid lines are the simulation output divided by $(\sin\alpha')^{1-\kappa}$ and the thick solid lines are the full simulation data. 
All curves are normalised by their maximal value on $\theta_{\uuEATS}$.
The thin lines match the analytic results completely. When accounting for the contribution of $\sin\alpha'$, the slope of the curve becomes flatter and as a result the center of the image becomes brighter with respect to the emission at $\theta\uuEATS$. The reason for that is that the average value of $(\sin\alpha')$ decreases with $\theta$. On the LOS $\unit{b}'\perp\unit{\theta'}$ and $\sin\alpha'=1$, while on $\theta\uuEATS$ $\unit{b}'$ can have an arbitrary direction and $\langle\sin\alpha'\rangle=\pi/4$. This effect is not captured by the analytic formula. 
In addition we can see the effect of the spectral slope on the curves, where the radio image is less limb brightened than the optical image. 
% In addition to the light curve shape, we also tested the distribution of the surface brightness across the EATS. We extracted the surface brightness from a snapshot of our simulation at two of the observed frequencies considered in this work. The surface brightness is averaged over the tangential direction to produce the dependency of it on $R_{\perp}$ alone. We find that the surface brightness is constant near the LOS and diverges as one nears $\theta_{\uuEATS}$. This is shown in fig. \ref{fig:SurfaceBrightnessTest} in solid lines where the blue line represents the optical at $\nu_{\rm obs}>\nu_m$ and the red line is the microwave frequency at $\nu_{\rm obs}<\nu_m$). Next, we compared our result to the analytical result of \cite{Sari1997Egg}, plotted in dashed lines in fig. \ref{fig:SurfaceBrightnessTest}. The growth rate of the surface brightness from our simulation (solid lines) is greater than that of the analytical model. Since all curves are normalized by their maximal value, they converge to each other as $R_{\perp}$ grows. We found the source of these differences to originate at our inclusion of the dependency of the direction of the magnetic field relative to the LOS. Our simulation converges to the analytical model when this factor of $(\sin\alpha')^{1-\kappa}$ is removed from it.

\begin{figure}
    \includegraphics[width=\columnwidth, trim=3cm 10cm 3cm 10cm]{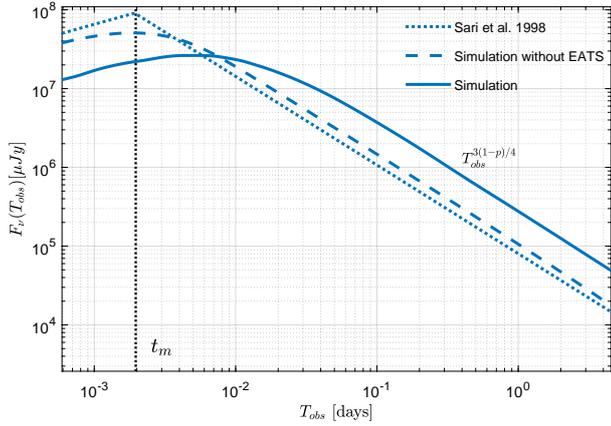}
    	\caption{A comparison of optical light curves from a spherical AG shock (no jet break) with a slow cooling emission. We present the analytic calculation from \citet[][dotted line]{Sari1998} together with a simulated light curves without accounting for light travel times (dashed line) and when accounting for it (solid line). The analytic light curve is multiplied by $8$ to match the simulated curve. See the text for an explanation on the origin of the factor. 
    	%a the slow cooling AG model considered in this work, without a jet break and with $q=0$ (on-axis jet). The dotted blue line corresponds to the "low frequency" light curve presented in \cite{Sari1998}. The dashed blue line is the light curve calculated using our model while omitting the EATS. The light curves produced by the simulation in this work, which includes the EATS, is plotted in solid blue lines. 
    	}
	\label{fig:FComp}
\end{figure}

%The calculations preformed in section \ref{sec:Emission} allow us to calculate the observed flux of the AG from our simulation. To test the correctness of our simulation, we compare the optical light curve from our simulation to the standard AG radiation model of \cite{Sari1998} in fig. \ref{fig:FComp}. We found that their calculation underestimates the observed flux by a factor of $\sim$8 due to approximations on the Doppler factor as well as averaging of the Lorentz factor of the areas contributing to the observed emission. \gb{I debated whether saying we overestimated the flux or they underestimated... which is better? } After correcting this factor, we found that our simulation, which treats the emission region as the spherical shock surface itself, and not the EATS, (dashed blue line) fits the calculation in \cite{Sari1998} (dotted blue line) well. In addition, we plot the light curve of our simulation which includes emission from the EATS. Most of the emission in this case will arrive from a ring with an angular radius $\theta_{\rm EATS}$ around the LOS and will be affected by the beaming effect. Since the angular time in relativistic systems is generally longer than the LOS time (\cite{Nakar2007}), the spectral break will also be delayed.
Next, we test the time evolution of the integrated flux in our code by comparing the lightcurve in the optical band to the analytic model of \citet{Sari1998}. The authors calculated the flux coming from a spherical shock, not accounting for photon arrival time effects and approximating the Doppler factor as a step function where $D=\Gamma$ for $\theta<1/\Gamma$ and $0$ elsewhere. 
To make a proper comparison we ran a limited version of the simulation with a spherical shock and without photon arrival times. 
The two lightcurves are presented in fig. \ref{fig:FComp}. The analytic curve is shown in dotted blue line and the result from the limited run in a dashed blue line. To match the results we multiply the analytic curve by a factor of $8$, which originates from two effects: 
i) The exact integral of  the Doppler factor over the spherical shock is 4 times larger than the integral of the step function. ii) another factor of 2 comes from differences in the constants used by \citet{Sari1998} when calculating the emission power, with respect to \citet{Rybicki1979}. 
To demonstrate the effect of the photon arrival time we add a plot from the full simulation (solid blue line). The peak of the lightcurve occurs at a later time, since the emission comes from a ring of matter with a larger Lorentz factor than on the LOS, thus the observed frequency in the proper frame is smaller and is crossed by $\nu\uum$ at a later time. 

\bsp	% typesetting comment
\label{lastpage}
\end{document}